\documentclass[aps,prl,reprint,groupedaddress,showpacs,amsmath,amssymb,twocolumn,floatfix]{revtex4-1}
\usepackage{graphicx}
\usepackage{bm}
\usepackage{color}
\usepackage[normalem]{ulem}
\usepackage{cleveref}
\usepackage{amsmath}

\newcommand{\wse}{WSe$_2$}

\newcommand{\WSe}{WSe$_2$}
\newcommand{\Rxx}{$R_{xx}$}
\newcommand{\Rxy}{$R_{xy}$}
\newcommand{\DRxy}{$\Delta R_{xy}$}
\newcommand{\DRxyn}{$\Delta R_{xy}^\nu$}
\newcommand{\DRxyb}{$\Delta R_{xy}^B$}
\newcommand{\EZV}{$E_Z^v$}
\newcommand{\EZS}{$E_Z^s$}

\newcommand{\Bperp}{$B_{\perp}$}
\newcommand{\Bpara}{$B_{\parallel}$}
\newcommand{\Bup}{$B_{\uparrow}$}
\newcommand{\Bdown}{$B_{\downarrow}$}
\newcommand{\Bupp}{$B^{\parallel}_{\uparrow}$}
\newcommand{\Bdownp}{$B^{\parallel}_{\downarrow}$}
\newcommand{\Bupperp}{$B^{\perp}_{\uparrow}$}
\newcommand{\Bdownperp}{$B^{\perp}_{\downarrow}$}
\newcommand{\densityunit}{$\times10^{12}$cm$^{-2}$}

\begin{document}

\title{Spin-orbit driven ferromagnetism at half moir\'e filling in magic-angle twisted bilayer graphene}




\author{Jiang-Xiazi Lin$^{1}$}
\author{Ya-Hui Zhang$^{2}$}
\author{Erin Morissette$^{1}$}
\author{Zhi Wang$^{1}$}
\author{Song Liu$^{3}$}
\author{Daniel Rhodes$^{3}$}
\author{K. Watanabe$^{4}$}
\author{T. Taniguchi$^{4}$}
\author{James Hone$^{3}$}
\author{J.I.A. Li$^{1}$}
\email{jia\_li@brown.edu}

\affiliation{$^{1}$Department of Physics, Brown University, Providence, RI 02912, USA}
\affiliation{$^{2}$ Department of Physics, Harvard University, Cambridge, MA 02138, USA}
\affiliation{$^{3}$Department of Mechanical Engineering, Columbia University, New York, NY 10027, USA}
\affiliation{$^{4}$National Institute for Materials Science, 1-1 Namiki, Tsukuba 305-0044, Japan}

\date{\today}

\maketitle

\textbf{Strong electron correlation ~\cite{Imada1998mott} and spin-orbit coupling (SOC) ~\cite{Hasan2010TI,Qi2011TI} provide two non-trivial threads to condensed matter physics. When these two strands of physics come together, a plethora of quantum phenomena with novel topological order have been predicted to emerge in the correlated SOC regime.
In this work, we examine the combined influence of electron correlation and SOC on a 2-dimensional (2D) electronic system at the atomic interface between magic-angle twisted bilayer graphene (tBLG) and a tungsten diselenide (\WSe) crystal. 
In such structure, strong electron correlation within the moir\'e flatband stabilizes correlated insulating states at both quarter and half filling, whereas SOC transforms these Mott-like insulators into ferromagnets, evidenced by robust anomalous Hall effect with hysteretic switching behavior.
The coupling between spin and valley degrees of freedom is unambiguously demonstrated as the magnetic order is shown to be tunable with an in-plane magnetic field, or a perpendicular electric field. In addition, we examine the influence of SOC on the isospin order and stability of superconductivity. Our findings establish an efficient experimental knob to engineer topological properties of moir\'e bands in twisted bilayer graphene and related systems. }

\begin{figure*}
\includegraphics[width=1\linewidth]{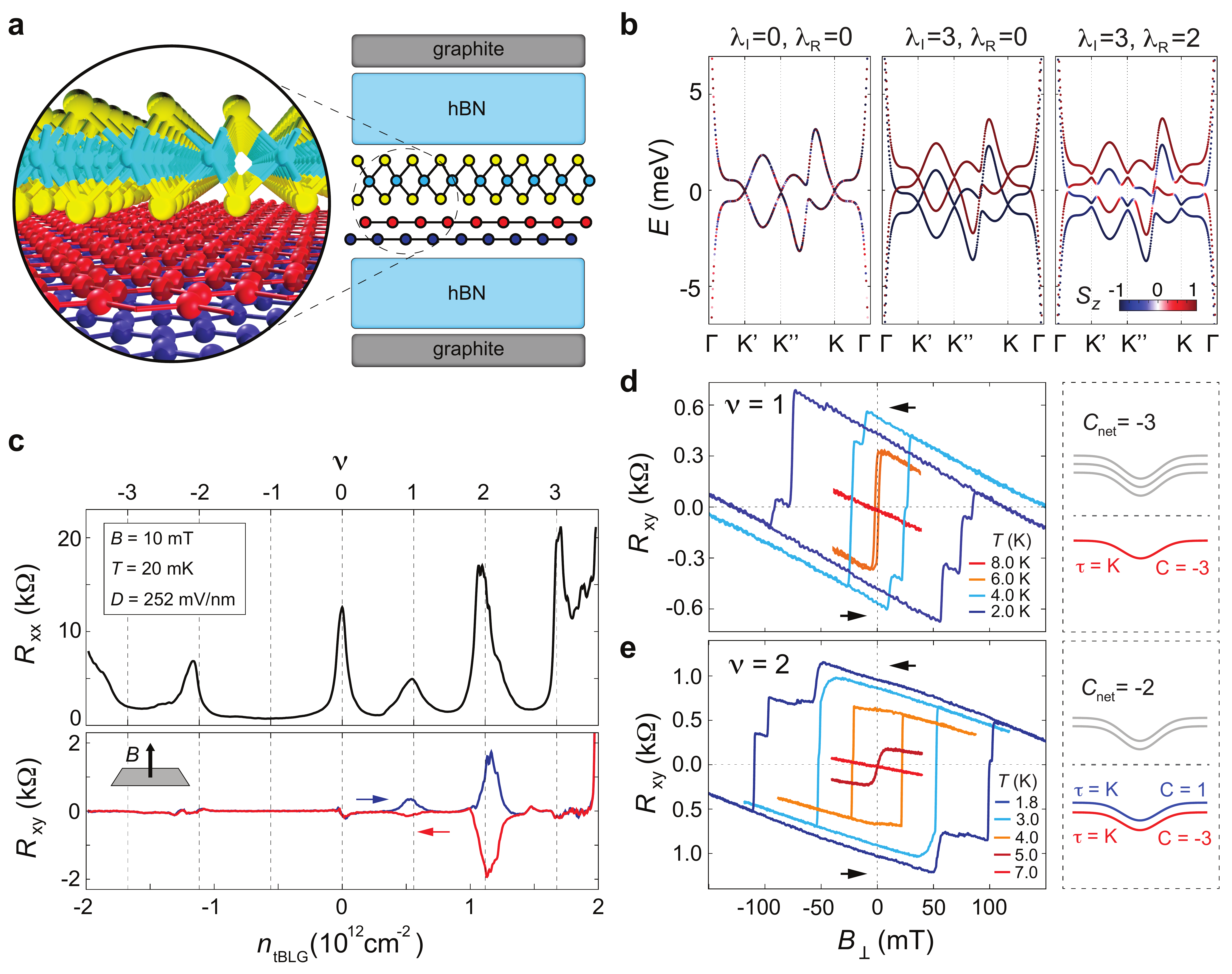}
\caption{\label{fig1} {\bf{Emerging ferromagnetic order from the tBLG/\wse\ interface}} (a) Schematic of the heterostructure consisting of a tBLG/\wse\ interface, which is doubly encapsulated with hBN and graphite. 
(b) Calculated dispersion of moir\'e bands for a single valley. $\lambda_I$ and $\lambda_R$ indicate the strength of Ising and Rashba SOC, respectively  in units of meV. Red and blue color denote the out-of-plane component of spin moment of each band. 
Chern number is expected to be zero for all energy bands in the absence of $\lambda_R$ (left and middle panel). The combination of strong Ising and Rashba SOC gives rise to non-zero valley Chern number.  
(c) Longitudinal and transverse resistance, $R_{xx}$ and $R_{xy}$, as carrier density $n_{tBLG}$ is swept back and forth. Carrier density $n_{tBLG}$ and moir\'e filling $\nu$ are denoted as the bottom and top axis, respectively. The measurement is performed at $B=10$ mT, $T = 20$ mK and $D = 252$ mV/nm.
(d-e) \Rxy\ measured at (d) $n_{tBLG} = 0.55$\densityunit\ near $\nu=1$ and (e) $n_{tBLG} = 1.22$\densityunit\ near $\nu=2$, as \Bperp\ is swept back and forth. These measurements are performed at $D = 0$. The hysteresis loop disappears at high temperature. Right panel: schematic band structure at (d) $\nu=+1$ and (e) $\nu=+2$. The two lowest bands feature the same valley index with non-zero Chern number of  $C = -3$ and $+1$. As a result, the ground state at $\nu= 2$ is valley polarized with net Chern number $C_{net}=-2$. }
\end{figure*}

\begin{figure*}
\includegraphics[width=0.75\linewidth]{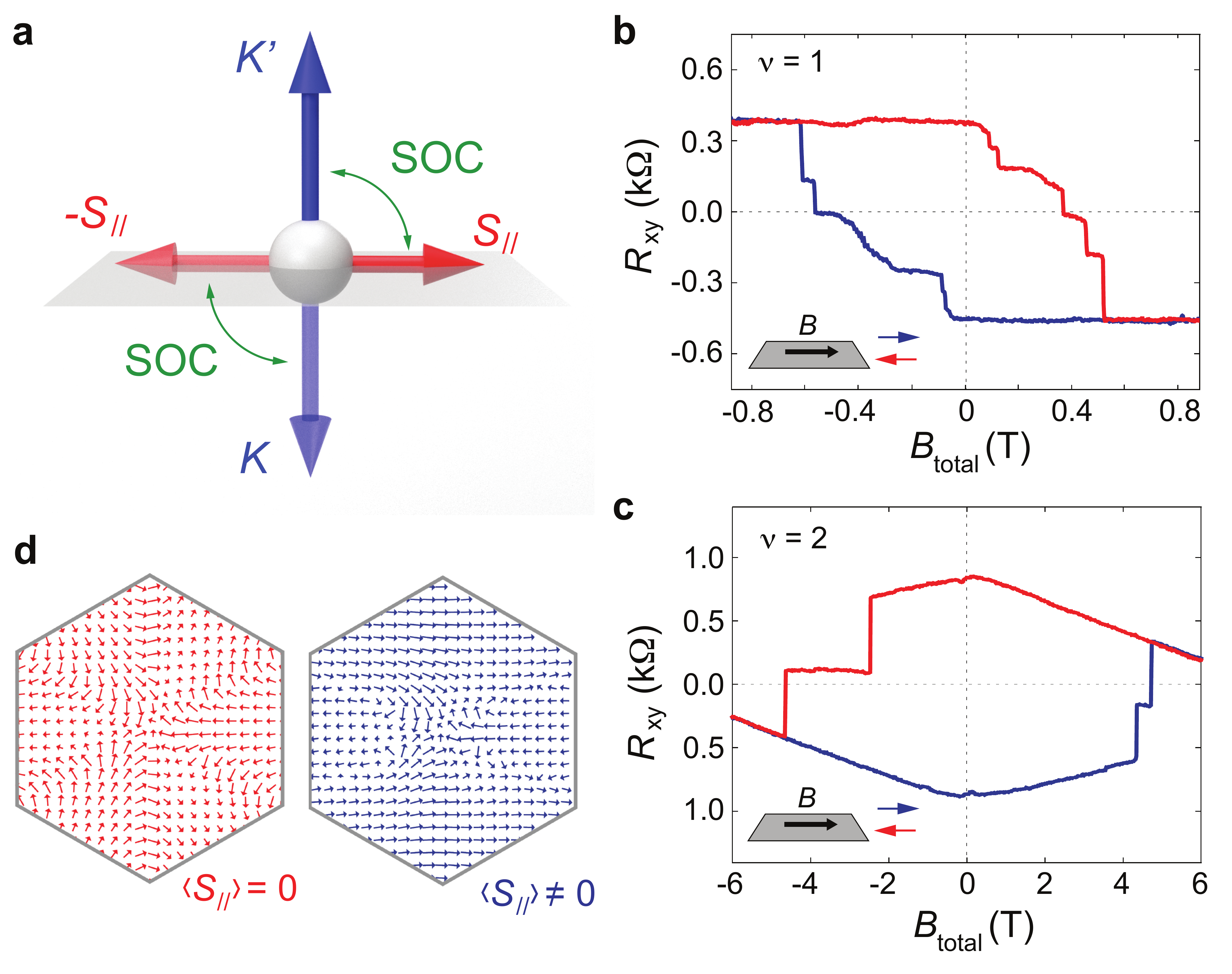}
\caption{\label{fig2} 
{\bf{Controlling magnetic order using an in-plane magnetic field}} (a) Schematic showing the effect of SOC, which couples the in-plane component of spin, $\pm S_{//}$,  with the out-of-plane component of valley, $\tau_z=K$ and $K'$.  (b-c) \Rxy\ as a function of in-plane $B$ field, which is aligned within $0.5^{\circ}$ of the tBLG/\WSe\ interface. Traces and retraces are shown as blue and red solid lines, respectively. The measurement is performed at (b) $\nu=+1$ and $D=-167$ mV/nm, (c) $\nu=+2$ and $D = 0$.  (d) The orientation of in-plane spin momentum over the MBZ for valley $K$, which is obtained by diagonalizing the single particle Hamiltonian (Fig.~1b, see SI for more detailed discussion) ~\cite{SI}. Here we show the first conduction band with valley index $K$ above the neutrality. In the presence of $C_3$ symmetry, $\langle \vec S_{\parallel}^{K} \rangle$ averages to zero  (left panel), whereas a uniaxial strain breaks $C_3$, resulting in non-zero $\langle \vec S_{\parallel}^{K} \rangle$ (right panel). 
}
\end{figure*}

The van der Waals (vdW) moir\'e structure has established an intriguing platform for exploring the interplay between correlation, topology and broken symmetry in 2-dimensional (2D) electronic systems. The rotational alignment between two sheets of vdW crystal gives rise to a flat moir\'e energy band where strong Coulomb correlation plays a dominating role in a rich landscape of emergent quantum phenomena
~\cite{Cao2018a,Cao2018b,Lu2019SC,Yankowitz2019SC,Chen2020ABC,Liu2020DBLG,Cao2020DBLG}. In a graphene moir\'e structure, breaking the $C_2T$ symmetry is shown to stabilize spontaneous orbital ferromagnetism at quarter and three-quarter filling, which is manifested in robust anomalous Hall effect (AHE) with hysteretic switching transitions  ~\cite{Sharpe2019,Serlin2019,Polshyn20201N2,Chen20201N2}.  
Unlike one and three-quarter filling, a potential orbital ferromagnetic state at half-filled moir\'e band would feature spin-unpolarized edge mode that is able to proximate superconducting pairing along a ferromagnet/superconducting interface ~\cite{Qi2010}. Such construction has been proposed to be key in realizing the majorana mode. However, an orbital ferromagnet is predicted to be energetically unfavorable in twisted graphene structures, owing to the inter-valley Hund's coupling ~\cite{ZhangYH2019Chern,Repellin2020AHE,Bultinck2020AHE,SI}. 



As an essential ingredient in forming certain topological phases, SOC adds an extra experimental knob to engineer the topological properties of moir\'e structures  ~\cite{Zhang2009TI,Hasan2010TI,Qi2011TI}. It was recently proposed that the introduction of SOC endows non-zero Berry curvature to the moir\'e energy band, making ferromagnetic order at half moir\'e filling a possibility without alignment with the hexgonal boron nitride (hBN) substrate ~\cite{Wang2020flat,SI}. 
Unlike bulk materials, where tuning the chemical composition is required to produce spin-orbit locking, vdW structures provide an alternative route through the proximity effect. Close proximity between graphene and transition metal dichalcogenide (TMD) crystals, such as \WSe, allows electron wavefunctions from both crystals to overlap and hybridize, endowing graphene with strong SOC ~\cite{Gmitra2015, Gmitra2017,Wang2015SOC,Wang2016SOC,Yang2017SOC,Avsar2014SOC,island2019spin,Arora2020SC}.
In this work, we use transport measurement to examine the effect of proximity-induced SOC on properties of the moir\'e band and its associated quantum phases.
 

The geometry of the tBLG/\WSe\ heterostructure is shown in Fig.~1a. An atomic interface is created by stacking a few-layer \WSe\ crystal on top of magic angle tBLG, which is further encapsulated with dual hBN and graphite crystals on top and bottom to achieve optimal sample quality ~\cite{Zibrov2018}.  
Transport measurement indicates excellent sample quality with low charge fluctuation  $\delta n \sim 0.08$ (10$^{12}$cm$^{-2}$) (see Fig.~\ref{fig:Rxy_at_CNP}). 
Longitudinal resistance $R_{xx}$ measured from tBLG exhibits a series of well-defined resistance peaks emerging at partial filling of the moir\'e band,  $\nu=-2$, $+1$, $+2$ and $+3$, which are associated with the correlated insulator (CIs) states. The positions of these peaks are consistent with a twist angle of $\theta \approx 0.98^{\circ}$. 
The tBLG and hBN substrate are maximally misaligned according to the optical image of the heterostructure (see Fig.~\ref{fig:crystal}) ~\cite{SI}, which is consistent with the fact that
the sample appears gapless at the CNP (see SI for more discussions regarding the coupling between tBLG and hBN) ~\cite{SI,Serlin2019,Sharpe2019}. 
Transverse resistance measurements reveal large Hall resistance at $\nu=+1$ and $+2$, which exhibits hysteretic switching behavior as the field-effect induced doping in tBLG, $n_{tBLG}$, is swept back and forth (Fig.~1c).  Hysteresis in magnetization reversal is also observed while sweeping an external magnetic field aligned perpendicular to the 2D interface, \Bperp (Fig.~1d-e). We note that the resistance peak at $\nu=+2$ vanishes at large in-plane $B$-field (see Fig.~\ref{fig:activation_gap}) ~\cite{SI}, indicating a spin-unpolarized isospin configuration. As such, the ground state is likely valley-polarized and the ferromagnetic order is orbital. This is further illustrated by a schematic representation of the band structure (right panels of Fig.~1d-e), where the two lowest conduction bands feature the same valley index with non-zero Chern number of $C=-3$ and $+1$. 

A valley polarized state at $\nu=2$  is unfavorable in the absence of SOC, due to the influence of inter-valley Hund's coupling (see SI for more detailed discussion) ~\cite{ZhangYH2019Chern,Repellin2020AHE,Bultinck2020AHE,SI}. As a result, observation of an orbital ferromagnet at half-filled moir\'e band has remained elusive ~\cite{Sharpe2019,Serlin2019,Stepanov2020}.
To this end, the AHE at $\nu=+2$ in our sample provides strong evidence that the moir\'e band structure is transformed by proximity-induced SOC, which is more dominant compared to inter-valley Hund's coupling ~\cite{Wang2020flat}.  Although the presence of SOC endows the moir\'e flatband with non-zero Chern number, as shown in Fig.~1d and e, the observed Hall resistance is much smaller than the expected value of quantum anomalous Hall effect. We ascribe this behavior to the presence of bulk conduction channels parallel to the chiral edge conduction, which could result from the presence of magnetic domain walls ~\cite{Sharpe2019} or sample disorder. A fully developed Chern insulator state could be realized in a sample with lower disorder or stronger SOC, hence a larger energy gap. 


To better understand the influence of proximity-induced SOC, we note that the introduction of SOC adds an extra term to the Hamiltonian:
\begin{equation}
h_{SOC}(\mathbf k) = \frac{1}{2}\lambda_{I} \tau_{z} s_{z} + \frac{1}{2}\lambda_{R} (\tau_{z} \sigma_{x} s_{y} - \sigma_{y} s_{x}).  \label{eq:soc}\\
\end{equation}
\noindent
Here $\tau$, $\sigma$ and $s$ denote Pauli matrix for valley, sublattice and spin  at each momentum $\mathbf k$, whereas $\lambda_I$ and $\lambda_R$ represent the Ising and Rashba SOC coefficients, respectively (See Eq.~S1-5 for more detailed discussions) ~\cite{SI}. 
The Ising SOC locks the valley moment $\tau_z$ with the spin moment $s_z$, whereas the Rashba term $\lambda_R$ locks the in-plane spin $s_x,s_y$ with the sublattice $\sigma_x,\sigma_y$ (the locking depends on the valley $\tau_z$). For tBLG without SOC, there is a $C_2$ symmetry defined as $C_2: \tau_x \sigma_x$ and time reversal symmetry defined as $T: i \tau_x s_y K$, where $K$ is the complex conjugation. Both the Ising and Rashba SOC break the $C_2$ symmetry, while preserving the time reversal. Therefore, $C_2 T$ is broken in the presence of proximity-induced SOC. 

The combination of Rashba SOC and time reversal symmetry gives rise to a valley-contrasting spin texture within the mini Brillouin Zone (MBZ) for each band: for any momentum $k$, spin for the two valleys points in opposite directions, $\vec s_K(\mathbf k)=-\vec s_{K'}(-\mathbf k)$. 
The average value of in-plane spin moment of each band is obtained by integrating over the MBZ for valley $K$ ($K'$), $\langle S_{K,K'}^{\parallel} \rangle=\frac{1}{N} \sum_{\mathbf k \in MBZ} \langle s_{K,K'}^{\parallel}(\mathbf k) \rangle$, where $N$ is the system size. We note that a broken $C_3$ rotation symmetry could lead to non-zero $\langle \vec S_{\parallel}^K \rangle$ and $\langle \vec S_{\parallel}^{K'} \rangle$. In this scenario, time reversal symmetry is preserved by the valley-contrasting spin texture $\langle \vec S_K^{\parallel} \rangle=-\langle \vec S_{K'}^{\parallel} \rangle$.
On the other hand, an orbital ferromagnetic state emerges because valley-polarized Chern bands are occupied, spontaneously breaking time reversal symmetry. Most remarkably, the combination of SOC, valley polarization and a broken $C_3$ rotation symmetry allows an in-plane magnetic field to couple to the orbital magnetic order through the in-plane Zeeman energy.

\begin{figure*}
\includegraphics[width=0.7\linewidth]{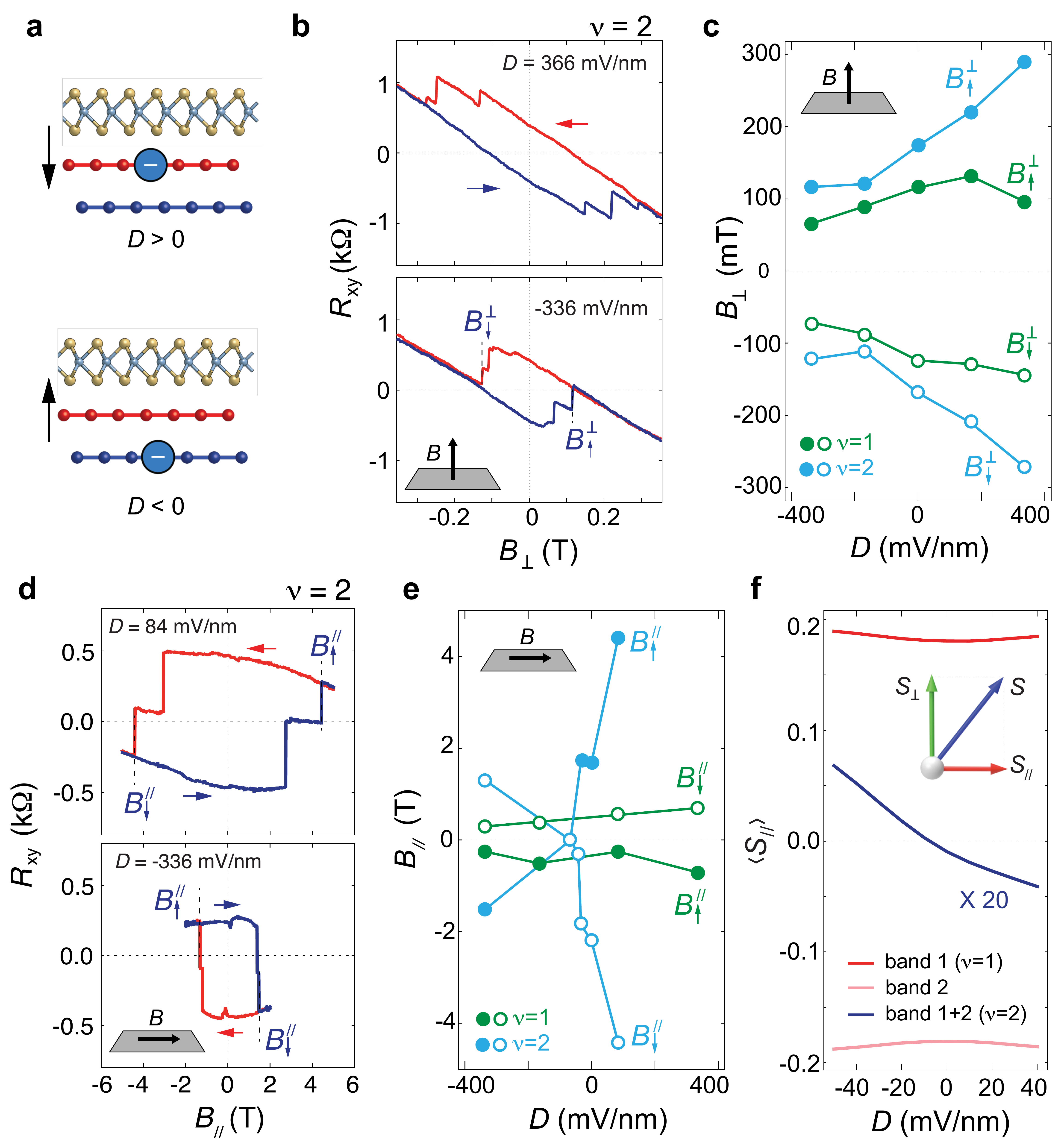}
\caption{\label{fig3} {\bf{Displacement-field dependence}}
(a) Schematic demonstrating the effect of an electric displacement $D$ on layer polarization. $B$-induced hysteresis loops of \Rxy\ measured at different $D$ with $B$-field aligned (b) perpendicular and (d) parallel to the 2D layers.    (c) Out-of-plane coercive field \Bupperp\ and \Bdownperp, and (e) in-plane coercive field \Bupp\ and \Bdownp for both $\nu=+1$ and $+2$ as a function of $D$. \Bup\ (\Bdown) is defined as the value of $B$ where the sign of \Rxy\ switches from negative to positive (positive to negative), whereas the superscript denotes the orientation of the $B$-field  (See Fig.~\ref{fig:Bloop}) ~\cite{SI}. Both in-plane and out-of-plane magnetic hysteresis behaviors for $\nu=+1$ are measured at $T = 20$ mK. At $\nu=+2$, out-of-plane magnetic hysteresis is measured at $T = 20$ mK, whereas in-plane hysteresis is measured at $T \leq 3$ K (see Fig.~\ref{fig:Bloop} for more details) ~\cite{SI}.
(f) $\langle S_{\parallel} \rangle$ calculated from the 4-band model as a function of $D$ for the two lowest energy bands, band 1 and 2 for the valley K. Both bands feature large $\langle S_{\parallel} \rangle$ that are mostly independent of $D$, shown as red and pink traces, respectively. The combination of band 1 and 2 yields a non-zero, albeit small, $\langle S_{\parallel} \rangle$, which changes sign around $D = 0$. 
}
\end{figure*}

\begin{figure*}
\includegraphics[width=0.95\linewidth]{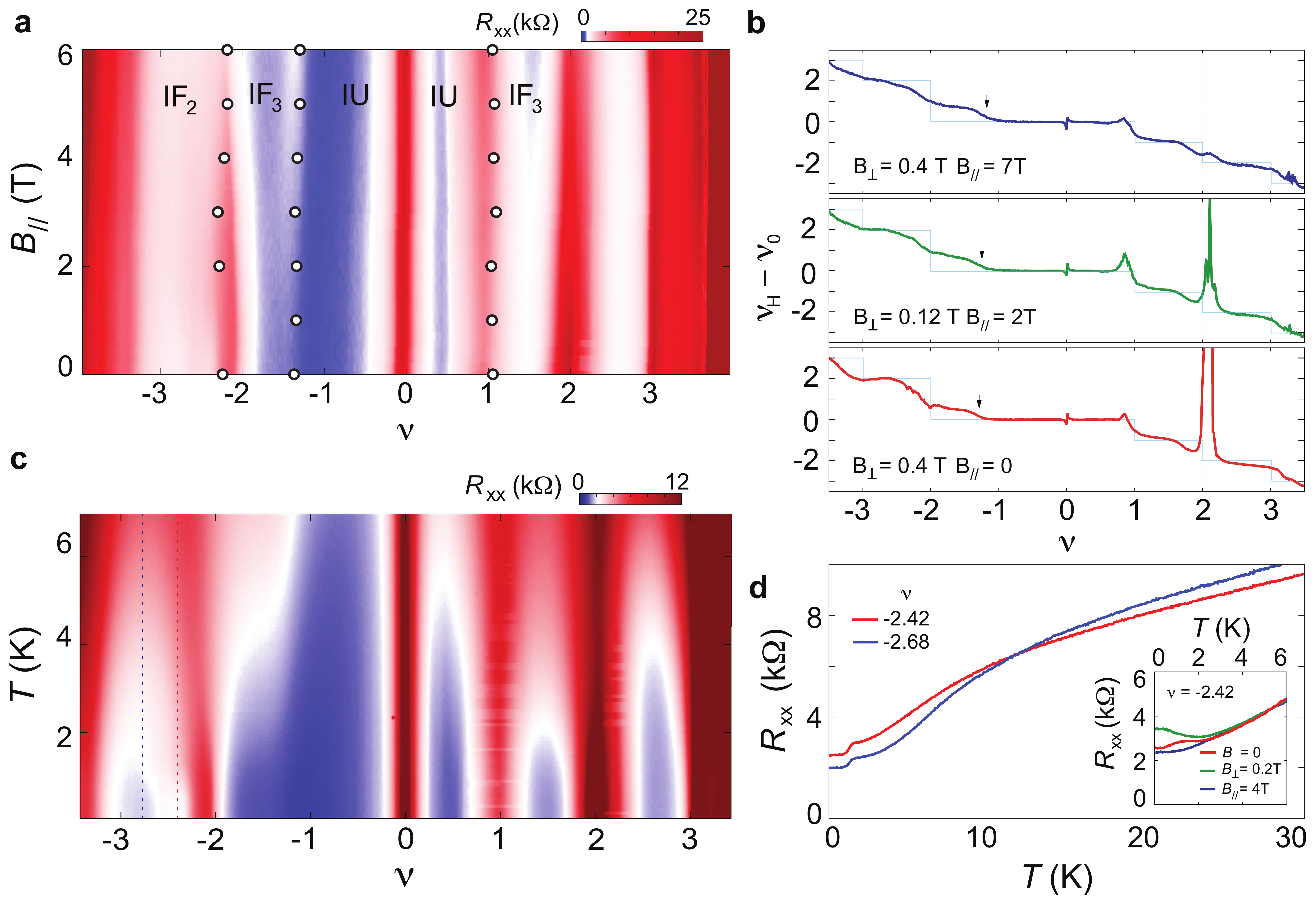}
\caption{\label{fig4} {\bf{Isospin order and the absence of superconductivity}} 
(a) $R_{xx}$ as a function of moir\'e filling $\nu$ and in-plane magnetic field \Bpara\ measured at $T=20$ mK. Carrier density is controlled by sweeping bottom gate voltage while top gate voltage is kept at zero. (b) Renormalized Hall density, $\nu_H-\nu_0$, expressed in electrons per superlattice unit cell, measured with different \Bpara\ and \Bperp\ at $D=252$~mV/nm. Circles in (a) denote the phase boundary between symmetry breaking isospin ferromagnets  (IF$_2$ and IF$_3$) and an isospin unpolarized state (IU), which are defined as the peak position in $d(\nu_H-\nu_0)/d\nu$. The expected Hall density steps of tBLG without SOC is shown as the light blue trace ~\cite{Saito2020pomeranchuk}.
(c) Longitudinal resistance $R_{xx}$ as a function of temperature and moir\'e filling measured at $B = 0$ and $D=252$~mV/nm. (d) $R-T$ line trace taken from (c) along vertical dashed lines. Inset, magnetoresistance displays no clear indication of Zeeman induced Cooper pair breaking: although $R_{xx}$ increases slightly with increasing \Bperp, it decreases with \Bpara.
}
\end{figure*}

This novel control on the out-of-plane magnetic order using an in-plane magnetic field \Bpara\ is demonstrated in Fig.~2b-c: as $B_{\parallel}$ is swept back and forth, \Rxy\ exhibits hysteretic switching behavior at both $\nu=+1$ and $+2$. Using Hall resistance measurement, we show that the out-of-plane component of the $B$-field is negligible compared to the out-of-plane coercive field, confirming that the magnetic order is indeed controlled by the in-plane component of the $B$-field, \Bpara\  (see Fig.~\ref{fig:Bperp}). At the same time,  direct coupling between the valley order and \Bpara\ is shown to be absent in tBLG samples without SOC ~\cite{Sharpe2021}, which rules out the orbital effect as a possible origin for the observed \Bpara-dependence. Taken together, we conclude that \Bperp\ and \Bpara\ couple to the magnetic order through different mechanisms: \Bperp\ directly controls the magnetic ground state  through valley Zeeman coupling, $E_Z^v=-\gamma_v B_{\perp} \tau_{z}$, whereas the influence of \Bpara\ arises from the combination of SOC and spin Zeeman coupling, $E_Z^s=-\gamma_s   \tau_z \vec B_{\parallel}\cdot \langle \vec S_{\parallel} \rangle$. Here $\tau_{z}$ corresponds to the valley polarization, $\gamma_v$ and $\gamma_s$ are valley and spin gyromagnetic ratio, respectively. 

We note a few intriguing properties of the $B$-induced hysteresis behavior in Fig.~1 and 2: (i) a large \Bpara\ stabilizes opposite magnetic orders at $\nu=+1$ and $+2$, which is evidenced by \Rxy\ with opposite signs (Fig.~2b-c). This indicates that $\langle S_{\parallel} \rangle$ points in opposite directions between $\nu=+1$ and $+2$. On the other hand, an out-of-plane B-field stabilizes the same magnetic order at different fillings (Fig.~1d-e), providing further confirmation that $\nu=+1$ and $+2$ features the same valley index and the ground state at half-filling is valley polarized; (ii) although the out-of-plane coercive fields are similar, the in-plane coercive field at $\nu=2$ is much bigger compared to $\nu=1$. This suggests that the average in-plane spin moment $\langle S_{\parallel} \rangle$ is much smaller at $\nu=2$, which is consistent with a predominantly spin unpolarized ground state (Fig.~\ref{fig:activation_gap}) ~\cite{SI}. 
It is worth pointing out that the observed behavior at $\nu=+2$ do not rule out alternative isospin configurations. Our results provide important constraints for future theoretical work to examine such possibilities.
The effective control of in-plane $B$-field on the magnetic order not only provides further validation that the orbital ferromagnetic order is stabilized by proximity-induced SOC, it also reveals that a broken $C_3$ rotational symmetry gives rise to non-zero $\langle S_{\parallel} \rangle$. 
A spontaneously broken $C_3$ symmetry naturally derives from the combination of strong SOC and nematic charge order ~\cite{Cao2020nematicity,Liu2018SDW,Sboychakov2019SDW}. Alternatively, a preferred in-plane direction for spin could result from small amount of uniaxial strain in the moir\'e lattice (Fig.~2d), which is shown to be common for tBLG samples ~\cite{Mcgilly2019seeing,Uri2019mapping}. 


The proximity-induced SOC arises from wavefunction overlap across the interface ~\cite{Gmitra2017}. The role of wavefunction overlap is recently demonstrated experimentally, as SOC strength is shown to depend on interlayer separation, which is tunable  with hydrostatic pressure ~\cite{Fulop2021}.
In the same vein, we show that SOC strength can be controlled with a perpendicular electric field $D$: under a positive (negative) $D$, charge carriers are polarized toward (away from) the \WSe\ crystal, resulting in increased (decreased) wavefunction overlap and stronger (weaker) SOC (Fig.~3a) ~\cite{Gmitra2017,Fulop2021}. Fig.~3 demonstrates the effect of $D$ by plotting the evolution of $B$-induced hysteresis loops: with increasing $D$, hysteresis loops exhibit larger coercive field (\Bupperp\ and \Bdownperp) for both $\nu=+1$ and $+2$ (Fig.~3b-c).  
Transport measurement near the charge neutrality point (CNP) shows that the width of the disorder regime remains the same over a wide range of $D$-field (see Fig.~\ref{fig:Rxy_at_CNP}b), suggesting that changes in the coercive field are not due to the influence of disorder across two graphene layers. Since the value of coercive field reflects the robustness of magnetic order, the $D$-dependence shown in Fig.~3c provides another indication that the orbital ferromagnetism is stabilized by proximity-induced SOC. 

Most interestingly, the effect of varying $D$ at $\nu=+2$ is drastically different in the presence of an in-plane magnetic field \Bpara: changing $D$ induces a magnetization reversal, which is evidenced by hysteresis loops with opposite signs in \Rxy\ (Fig.~3d and Fig.~\ref{fig:D}a) ~\cite{SI}. 
The $D$-induced reversal is also manifested in the sign change of coercive field \Bupp\ and \Bdownp\ at $D \sim -100$ mV/nm (Fig.~3e).
A possible explanation for this unique $D$-dependence is obtained by examining the average in-plane spin moment for valley $K$, $\langle \vec S_K^{\parallel} \rangle$. Calculation using the 4-band model shows that $\langle \vec S_K^{\parallel} \rangle$ changes sign with varying $D$  at $\nu=2$ (Fig.~3f), indicating that an in-plane $B$-field favors opposite valley polarizations at different $D$.
In comparison, Fig.~3f shows that no sign change in $\langle \vec S_K^{\parallel} \rangle$ is expected at $\nu=1$, which is consistent with our observation (Fig.~3e). 
Since carrier density remains the same when changing $D$, the $D$-controlled magnetization reversal represents an unprecedented electric field control on the magnetic order, which is made possible by proximity induced SOC.

Next, we turn our attention to the effect of SOC on the isospin polarization and superconductivity. 
In the absence of SOC, the ground state at $\nu=-1$ is shown to be isospin unpolarized at $B=0$. The application of a large in-plane magnetic field lifts the isospin degeneracy, stabilizing  an isospin ferromagnetic state ($IF_3$) near $\nu=-1$, which is separated from the unpolarized state  (IU) by a resistance peak in \Rxx\ and a step in the Hall density ~\cite{Saito2020pomeranchuk,Rozen2020pomeranchuk}. 
Fig.~4a-b show that this phase boundary between $IF_3$ and $IU$ extend to \Bpara $=0$ in the presence of proximity-induced SOC. Taken together, the influence of SOC on the iso-spin degeneracy is comparable to a large in-plane magnetic field (see Fig.~\ref{fig:compare}) ~\cite{SI}.
Most remarkably, we show that the superconducting phase is unstable against proximity-induced SOC in our sample. As shown in Fig.~4c, no zero-resistance state is observed over the full density range of the moir\'e band. In addition, $R_{xx}-T$ traces exhibit no clear downturn with decreasing $T$ at moir\'e filling $\nu=-2-\delta$, where a robust superconducting phase usually emerges in magic-angle tBLG (Fig.~4d). Since strong Ising and Rashba SOC break $C_2T$ symmetry, the absence of superconductivity, combined with the emergence of AHE are potentially consistent with a recent theoretical proposal that $C_2T$ symmetry is essential for stabilizing the superconducting phase in tBLG ~\cite{Khalaf2020C2T}.
 


It is worth pointing out that our results are distinct from another experimental report showing  robust superconductivity stabilized by SOC in tBLG away from the magic angle ~\cite{Arora2020SC}. 
Apart from the difference in twist angle range, these distinct observations could result from a few factors that will be discussed in the following.
The $D$-dependence shown in Fig.~3 suggests that a dual-gated geometry, which allows independent control on $D$ and $n_{tBLG}$, is key to investigating the influence of proximity-induced SOC in \WSe/tBLG samples. When carrier density is controlled with only the bottom gate electrode ~\cite{Arora2020SC}, doping tBLG with electron also gives rise to a $D$-field that pulls electrons away from the \WSe\ crystal (see Fig.~\ref{FIG.fab}) ~\cite{SI}. This results in weaker SOC strength, which could contribute to the observation of superconductivity in singly-gated tBLG/\WSe\ samples ~\cite{Arora2020SC}. 
In addition, it is proposed that the strength of proximity-induced SOC is sensitive to the rotational alignment between graphene and \WSe: strong SOC is expected when graphene is rotationally misaligned with \WSe\ by $10-20^{\circ}$, whereas perfect alignment produces weak proximity-induced SOC ~\cite{Koshino2019SOC}. If confirmed, this rotational degree of freedom could provide an additional experimental knob to engineer moir\'e band structure ~\cite{Kennes2021moire}.  Our sample features a twist angle of $\sim$ $16^{\circ}$ (Fig.\ref{FIG.fab}f) between tBLG and \WSe, falling in the range that is predicted to induce the strongest SOC strength. The effect of rotational misalignment between tBLG and \WSe\ is investigated in two additional samples near the magic angle: AHE and hysteresis loops are observed at $\nu=+2$ in sample A1 where tBLG and \WSe\ are misaligned at $\sim 10^{\circ}$ (Fig.~\ref{fig:erin}). On the other hand, tBLG and \WSe\ are perfectly aligned in sample A2, where AHE is absent  (Fig.~\ref{fig:zhi}) ~\cite{SI}. The superconducting phase is absent or significantly suppressed in all samples. These observations provide experimental support for the notion that the rotational misalignment between tBLG/\WSe, thus the SOC strength, play a key role in determining the stability of the ferromagnetic and superconducting states. 
Although transport measurement alone cannot definitively confirm the influence of SOC on the superconducting phase near the magic angle, our results could motivate future efforts, both theoretical and experimental, to investigate the influence of SOC on moir\'e structures as a function of graphene twist angle and graphene/\WSe\ misalignment ~\cite{Kennes2021moire}.

\section*{Acknowledgments}
The authors acknowledge helpful discussions with A.F. Young, M. Yankowtiz and A. Vishwanath, as well as experimental assistance from A. Mounce and M. Lilly. This work was primarily supported by Brown University. Device fabrication was performed in the Institute for Molecular and Nanoscale Innovation at Brown University. This work was performed, in part, at the Center for Integrated Nanotechnologies, an Office of Science User Facility operated for the U.S. Department of Energy (DOE) Office of Science. J.L., E.M. AND J.I.A.L. acknowledge the use of equipment funded by the MRI award DMR-1827453. S.L., D.R. and J.H. acknowledge support from MRSEC on Precision-Assembled Quantum Materials (PAQM) - DMR-2011738.  K.W. and T.T. acknowledge support from the EMEXT Element Strategy Initiative to Form Core Research Center, Grant Number JPMXP0112101001 and the CREST(JPMJCR15F3), JST.

\section*{Competing financial interests}
The authors declare no competing financial interests.

\newpage

\newpage
\clearpage

\pagebreak
\begin{widetext}
\section{Supplementary Materials}

\begin{center}
\textbf{\large Proximity-induced spin-orbit coupling and ferromagnetism in magic-angle twisted bilayer graphene}\\
\vspace{10pt}
Jiang-Xiazi Lin,
Ya-Hui Zhang, Erin Morissette, Zhi Wang,  Song Liu, Daniel Rhodes, K. Watanabe, T. Taniguchi, James Hone, J.I.A. Li$^{\dag}$\\ 
\vspace{10pt}
$^{\dag}$ Corresponding author. Email: jia$\_$li@brown.edu
\end{center}

\noindent\textbf{This PDF file includes:}

\noindent{Supplementary Text}

\noindent{Materials and Methods}

\noindent{Figs. S1 to S19}

\noindent{References (44-47)}

\renewcommand{\thefigure}{S\arabic{figure}}
\setcounter{figure}{0}
\setcounter{equation}{0}

\newpage







\section{Supplementary Text: band structure calculation and theoretical analysis}

\subsection{Band structure from continuum model}

First, let us consider the pure twisted bilayer graphene without any spin-orbit coupling. We calculate the band structure  using the standard continuum model. The two valleys K and K' will form separate bands in the mini Brillouin zone (MBZ), and they are related by time reversal symmetry. Hence it is sufficient to focus only on the valley K. The Hamiltonian is
\begin{equation}
	H=H_{top}+H_{bottom}+H_M
\end{equation}

We have  
\begin{equation}
	H_{top}=\sum_{\mathbf k}(c^\dagger_{A}(\mathbf k), c^\dagger_{B}(\mathbf k))\left(
	\begin{array}{cc}
	0& -\frac{\sqrt{3}}{2}t(\tilde k_x-i \tilde k_y)\\
	-\frac{\sqrt{3}}{2}t(\tilde k_x+i \tilde k_y) &0
	\end{array}\right)  \left(\begin{array}{c}  c_A(\mathbf k)\\  c_B(\mathbf k)\end{array}\right)
\end{equation}
where $\mathbf{\tilde k}=R(-\theta/2) \mathbf {k}$, with twist angle $\theta$. $R(\varphi)$ is the transformation matrix for anticlockwise rotation with angle $\varphi$.

and 

\begin{equation}
	H_{bottom}=\sum_{\mathbf k}(\tilde c^\dagger_{A}(\mathbf k), \tilde c^\dagger_{B}(\mathbf k))\left(
	\begin{array}{cc}
    0& -\frac{\sqrt{3}}{2}t(\tilde k_x-i \tilde k_y)\\
	-\frac{\sqrt{3}}{2}t(\tilde k_x+i \tilde k_y) &0 
	\end{array}\right)  \left(\begin{array}{c} \tilde c_A(\mathbf k)\\ \tilde c_B(\mathbf k)\end{array}\right)
\end{equation}
where $\mathbf{\tilde k}=R(\theta/2) \mathbf {k}$. In the above, $k$ is in the unit of $\frac{1}{a}$, where $a=0.246$ nm is the graphene lattice constant.

Finally the interlayer moir\'e tunneling term is
\begin{equation}
	H_M=\sum_{\mathbf k} \sum_{j=0,1,2} (\tilde c^\dagger_{A}(\mathbf k), \tilde c^\dagger_{B}(\mathbf k))\left(
	\begin{array}{cc}
	\alpha t_M& t_M e^{-i  \frac{2\pi}{3} j}\\
	t_M e^{i  \frac{2\pi}{3} j} &\alpha t_M 
	\end{array}\right)  \left(\begin{array}{c} c_{A_1}(\mathbf k+\mathbf{Q}_j)\\ c_{B_1}(\mathbf k+\mathbf{Q}_j)\end{array}\right)+h.c.
\end{equation}
where $\mathbf Q_0=(0,0)$, $\mathbf Q_1=\frac{1}{a_M}(-\frac{2\pi}{\sqrt{3}},-2\pi)$ and $\mathbf Q_2=\frac{1}{a_M}(\frac{2\pi}{\sqrt{3}},-2\pi)$, with $a_M=\frac{1}{2 \sin \frac{\theta}{2}}$ as the moir\'e lattice constant.

We use parameters $t=2380$ meV and $t_M=110$ meV and $\alpha=0.8$.  To model strain, we simply replace the $t_M$ of the $\mathbf Q_0$ component  with $(1-\beta)t_M$, where $\beta$ is proportional to the strength of the strain. We use $\beta=0.02$ for Fig.2(d).

The spin-orbit coupling is added to the top layer by modifying $H_{top}(\mathbf k)\rightarrow H_{top}(\mathbf k)+h_{SOC}(\mathbf k)$, where

\begin{align}
    h_{SOC}(\mathbf k)=\frac{1}{2}\lambda_I \tau_z s_z+\frac{1}{2} \lambda_R(\tau_z \sigma_x s_y-\sigma_y s_x)+\lambda_K\tau_z \sigma_z s_z
\end{align}
where $\tau_\mu$, $\sigma_\mu$ and $s_\mu$ with $\mu=0,x,y,z$ label the Pauli matrix for valley, sublattice and spin space respectively. $\lambda_i,\lambda_R,\lambda_K$ are Ising, Rashba and Kane-Mele spin-orbit couplings respectively.  We set $\lambda_K=0$ as it is known to be small in graphene.   Proximity effect from TMD layer provides $\lambda_I$ and $\lambda_R$.

Because the spin-orbit coupling only involves $\tau_z$, it is diagonal in the valley space and we can still only focus on valley $K$.  The other valley $K'$ is related by time reversal symmetry $T:i\tau_x s_y$, which flips both the valley and spin.  Without SOC, there is an inversion symmetry: $C_2: \tau_x \sigma_x$.  We note that when $\lambda_I\neq 0$ and $\lambda_R \neq 0$, the combined symmetry $C_2 T$ is broken and there is no symmetry to protect the gapless Dirac point in the TBG without SOC.  The breaking of $C_2 T$ symmetry is crucial to have non-zero Berry curvature and the observed anomalous Hall effect.

\subsection{Effect of in-plane magnetic field}
We discuss the coupling to the in-plane magnetic field $\vec B_\parallel$ here. Just from the symmetry, the coupling of valley to the magnetic field within each band is constrained in the form:

\begin{equation}
    H'=-\frac{1}{2}\sum_{\mathbf k} g_{a;z}(\mathbf k) B_z-\frac{1}{2}\sum_{\mathbf k} \vec{g}_{a;\parallel}(\mathbf k)\cdot \vec{B}_\parallel 
\end{equation}
where $a=K,K'$ labels the valley index. $\vec{g}_a(\mathbf k)$ is the effective valley Zeeman coupling at momentum $\mathbf k$. Time reversal symmetry always constrains $\vec{g}_{K}(\mathbf k)=-\vec{g}_{K'}(-\mathbf k)$.  If there is $C_3$ symmetry, there is another constraint that $\vec{g}_\parallel(C_3 \mathbf k)=C_3 \vec{g}_\parallel(\mathbf k)$, which leads to $\langle \vec g_\parallel \rangle=\frac{1}{N} \sum_{\mathbf k \in MBZ} \vec{g}_{\parallel}(\mathbf k) =0$. 

We assume the anomalous Hall state observed at $\nu_T=1,2$ is valley polarized. Therefore, the energy difference for the oppositely valley polarized state under the magnetic field  is decided by  the averaged value of the g factor:
\begin{equation}
    \Delta E=-  g_z B_z - \vec g_\parallel \cdot \vec B_\parallel
\end{equation}
with
\begin{equation}
    \vec{g}=\frac{1}{N} \sum_\mathbf k\vec{g}_{K}(\mathbf k)
\end{equation}

In the above we have used the fact that $\vec{g}_{K}(\mathbf k)=-\vec{g}_{K'}(-\mathbf k)$. The expression also holds only at linear order of magnetic field.  

If there is $C_3$ symmetry, then $\vec g_\parallel=0$ and the coupling to the in-plane magnetic field vanishes at the linear order. In the experiment we find that in-plane magnetic field can switch the anomalous Hall effect, suggesting that $\vec g_\parallel \neq 0$ and therefore $C_3$ rotation symmetry is broken. This conclusion does not depend on detailed mechanism of the coupling.

Next we turn to the microscopic origin of the effective coupling. For the coupling to the out of plane magnetic field, we have:
\begin{equation}
    g_{a;z}(\mathbf k)=m_a(\mathbf k)+g_s\mu_B s_{a;z}(\mathbf k)
\end{equation}
where  $g_s\approx 2$ is the spin g factor. $s_{a;z}(\mathbf k)$ is the spin polarization at momentum $\mathbf k$ for the valley $a=K,K'$. The first term $m_a(\mathbf k)$ is the orbital magnetic moment for valley $a=K,K'$. Note that the second term arises because the two valley have opposite $s_z$ due to the Ising SOC. In the system without SOC, the two valley can have the same spin polarization and there is no contribution from the second term.   However, we believe the first orbital magnetic term is the dominant one as the associated valley g factor was usually several times larger than the spin g factor ~\cite{ZhangYH2019AHE}.

The coupling to the in-plane field can be similarly divided to the spin and orbital parts:
\begin{equation}
    \vec g_{a;\parallel}(\mathbf k)=g_{o;a;\parallel}(\mathbf k)+g_s \mu_B \vec s_{a;\parallel}(\mathbf k)
\end{equation}
where the first term arises because of the shifting of the Dirac cones of the two layers ~\cite{Cao2020nematicity}. The second term exists because the Rashba SOC provides valley-contrasting in-plane spin component.  What matters is the averaged value in the momentum space. First, let us ignore the orbital part and focus on the spin Zeeman part.  When the $C_3$ rotation symmetry is broken, we find a non-zero $\langle \vec s_{\parallel} \rangle$ and therefore we can have $\vec g_\parallel \neq 0$. For filling $\nu_T=2$, we need two bands within the same valley to be both occupied. In this case, the net averaged value of $\langle \vec s_{\parallel} \rangle$ is almost cancelled and the coupling $\vec g_\parallel$ after average should be small as we have seen experimentally. Note that the direction of  $\vec s_\parallel$ is fixed  by $C_3$ breaking pattern and there is then a $\cos \theta$ dependence for the coupling strength of in-plane magnetic field along different directions.

The  orbital part $g_{o;\parallel}$ may also be non-zero when $C_3$ is broken. If $C_3$ breaking is due to strain, our estimation shows that $\vec g_{o;\parallel}$ is comparable to the spin Zeeman coupling contribution $g_s \mu_B \vec s_\parallel$. This is mainly because $\vec s_\parallel$ can be only $0.1$ with a $2\%$ strain. In this case, it is hard to isolate the contribution from the orbital and the spin Zeeman coupling. However, if $C_3$ is broken by strong correlation effect (for example, the in-plane spin is polarized due to interaction effects), the in-plane spin polarization $\vec s_\parallel$ may be significantly enhanced, making the spin Zeeman coupling part the dominant contribution.  In our calculation under strain, $\vec g_{o;\parallel}$ for the two bands within the same valley have the same sign, in contrast to the spin part $g_s \mu_B \vec s_\parallel$ which gets cancelled  when both bands are occupied.  Therefore, our observation of a  much smaller in-plane magnetic field coupling at $\nu_T=2$ than at $\nu_T=1$  is consistent with the scenario that the spin Zeeman coupling is the dominant contribution at $\nu_T=1$. This requires a large in-plane spin polarization for $\nu_T=1$. In future, it is interesting to measure $s_\parallel$ directly to quantify the magnitude of the spin Zeeman coupling contribution. For $\nu_T=2$, we believe the spin polarization is small, resuling in a much smaller coupling $g_\parallel$.

\subsection{Theoretical analysis of the anomalous Hall effect}

With non-zero Rashba SOC, $C_2T$ symmetry is broken and now we have non-zero Berry curvature and each band has non-zero valley Chern number. The total Chern number of the two valleys cancel each other. However, under strong correlation, the valley can be polarized through the quantum Hall ferromagnetism (QHFM) mechanism. In this case there can be anomalous Hall effect.   Anomalous Hall effect and orbital ferromagnetism have been observed in various moir\'e  systems at filling $\nu=1$. The observation of an anomalous Hall effect at $\nu=2$ here is a new result, which deserves a separate theoretical analysis.

In the usual graphene moire system without spin-orbit coupling, there are three almost degenerate QHFM states: (I) Spin polarized, valley unpolarized state: the bands $K,\uparrow$, $K',\uparrow$ are occupied. (II) Valley unpolarized, spin unpolarized state: the bands $K,\uparrow$, $K'\downarrow$ are occupied. (III) Valley polarized, spin unpolarized state: the bands $K,\uparrow$,$K\downarrow$ are occupied.  A small term is needed to select from the above three states. Without SOC, an inter-valley Hund's coupling dominates and selects the state (I) or (II) depending on its sign. In any case there is no anomalous Hall effect.

In the current system with SOC, the selection among these three states needs to be modified. In the following, we show that the Rashba SOC can favor the valley polarized state even at $\nu=2$.

Let us focus on the conduction band, labelling the two conduction bands for each valley as $a=1,2$. We use $\alpha=K,K'$ to label the two valleys and then each band is specified by the combined index $(\alpha,a)$. With Rashba SOC, the direction of the spin depends on the momentum $\mathbf k$ within each band.  There are two states which are favored in the strong interaction regime: (I) We occupy the band $K,1$ and $K',1$.  (II) We occupy the band $K,1$ and $K,2$.  The state (I) does not support anomalous Hall effect, while the state (II) polarizes the valley and supports anomalous Hall effect.  Next we discuss the competition between these two states.

The Hartree-Fock energy of the state I is:

\begin{equation}
    E_1=2\sum_{\mathbf k}\xi_1(\mathbf k)-2\frac{1}{N}\sum_{\mathbf k, \mathbf q}V(\mathbf q) |\Lambda_1(\mathbf k,\mathbf k+\mathbf q)|^2 +H_{Hartree}
\end{equation}
where $\xi_a(\mathbf k)$ is the dispersion of the band $a=1,2$ for valley $K$.  $\Lambda_a(\mathbf k,\mathbf k+\mathbf q)=<\mu_{a,K}(\mathbf k+q)|\mu_{a,K}(\mathbf k)>$ is the form factor obtained when projecting the density operator in the narrow conduction bands. In the above, the second term is the Fock energy and the third term is the Hartree energy. $N$ is the number of  moir\'e sites in the system. $H_{Hartree}$ is the Hartree energy.  The Hartree energy is suppressed because  the Form factor $\Lambda(\mathbf k,\mathbf k+\mathbf G)$ is small for large moir\'e  reciprocal vector $\mathbf G$. Here we only keep the Fock energy, which is the most essential term to cause QHFM.  The factor $2$ in the Fock energy comes from the two valleys.

Similarly, the energy of the state II is

\begin{equation}
    E_1=\sum_{\mathbf k}(\xi_1(\mathbf k)+\xi_2(\mathbf k)-\frac{1}{N}\sum_{\mathbf k, \mathbf q}V(\mathbf q) |\Lambda_1(\mathbf k,\mathbf k+\mathbf q)|^2-\frac{1}{N}\sum_{\mathbf k, \mathbf q}V(\mathbf q) |\Lambda_2(\mathbf k,\mathbf k+\mathbf q)|^2-2 \frac{1}{N}\sum_{\mathbf k, \mathbf q}V(\mathbf q) |\Lambda_{12}(\mathbf k,\mathbf k+\mathbf q)|^2  +H_{Hartree}
\end{equation}
where $\Lambda_{12}(\mathbf k,\mathbf k+q)=<\mu_{1,K}(\mathbf k+q)|\mu_{2,K}(\mathbf k)>$ is the form factor relating the two conduction bands of the valley K. Note that the spin of the two bands are now not necessarily orthogonal and generically $\Lambda_{12}(\mathbf k,\mathbf k+q)$ is not zero.  This is a new term which only exists because of the SOC.

If we assume that $\frac{1}{N}\sum_{\mathbf k, \mathbf q}V(\mathbf q) |\Lambda_2(\mathbf k,\mathbf k+\mathbf q)|^2 \approx \frac{1}{N}\sum_{\mathbf k, \mathbf q}V(\mathbf q) |\Lambda_1(\mathbf k,\mathbf k+\mathbf q)|^2$ and ignore the Hartree energy, the energy difference of the two states is:

\begin{equation}
    E_1-E_2 \sim \sum_{\mathbf k}(\xi_2(\mathbf k)-\xi_1(\mathbf k))-2 \frac{1}{N}\sum_{\mathbf k, \mathbf q}V(\mathbf q) |\Lambda_{12}(\mathbf k,\mathbf k+\mathbf q)|^2
\end{equation}
The first term is always positive, but the second term is negative and proportional to the energy scale of the Coulomb interaction.  Given that Coulumb interaction is significantly larger than the kinetic energy, it is possible that $E_1-E_2<0$ and the state II is favored.

We note that the valley polarized state is favored because of an extra term in the Fock energy.  This can be intuitively summarized as a generalized Hund's rule: When one band is occupied, then it is energetically favorable to occupy another band which is connected to the occupied band by the density operator, i.e. their spin and valley components are not orthogonal.  In the previous system without SOC, if we occupy the band $(K,\uparrow)$, then the occupation of the next band is arbitrary because the other bands are all orthogonal to the occupied band and can not be connected by the density operator (if we ignore the small inter-valley component).  In contrast, in the current system with SOC, if we occupy the band $(K,1)$,   then the band $(K,2)$ is connected with the occupied band because their spin component can have non-zero overlaps. On the other hand the band $(K',1)$ is not connected to the occupied band $(K,1)$ because the valley components are orthogonal.  Therefore it is favorable to occupy the band $(K,2)$ if the band $(K,1)$ is occupied first.  Hence the favored state is to occupy $(K,1),(K,2)$ (or its time reversal partner $(K',1),(K',2)$).


\section{Materials and Methods}

\subsection{Device Fabrication}

\begin{figure*}
\includegraphics[width=0.85\linewidth]{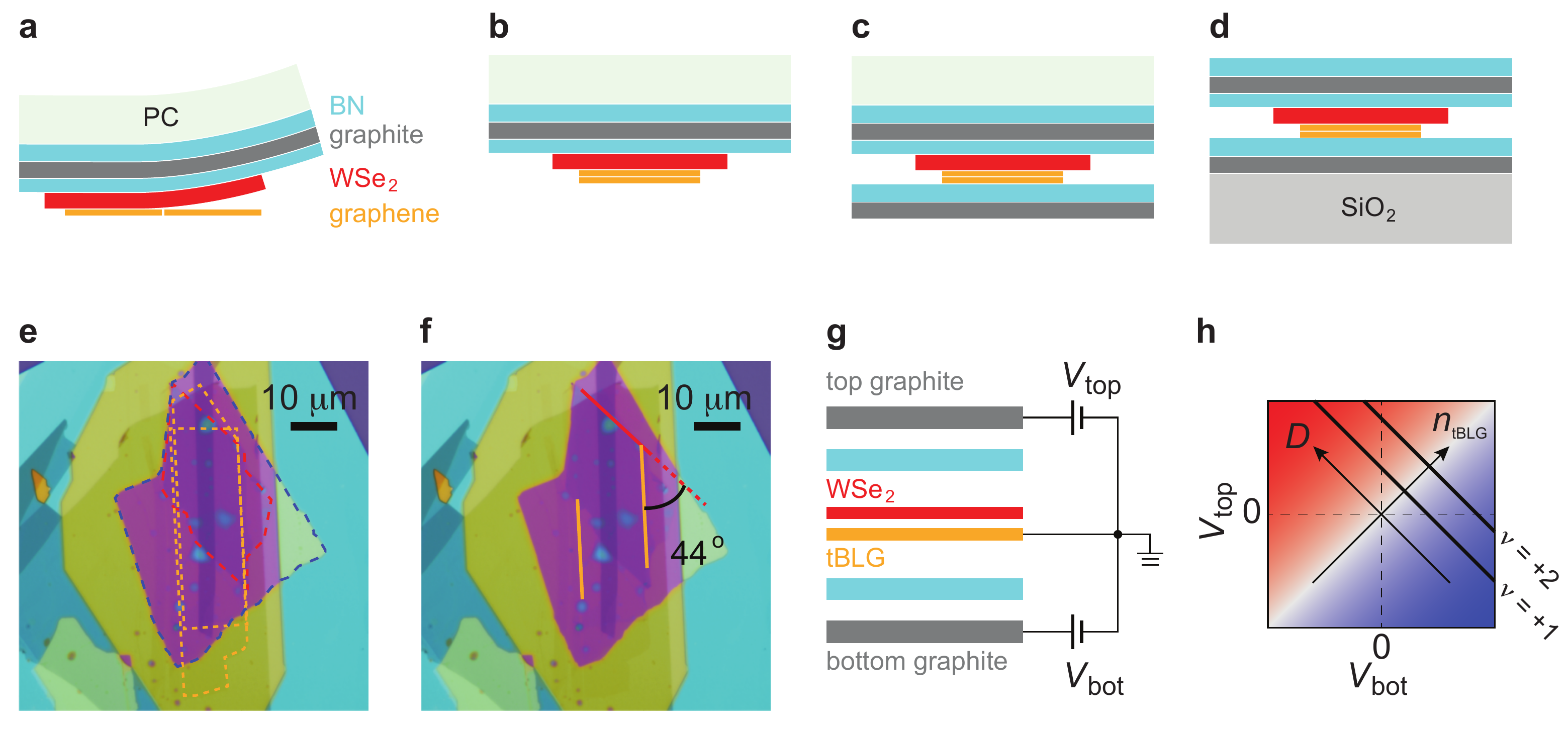}
\caption{\label{FIG.fab} {\bf{Device fabrication and measurement}} (a-d) Schematics illustrating the vdW assembly procedures for creating the tBLG/\WSe\ heterostructure. Each layer of the two-dimensional material is exfoliated onto a silicon chip, which is then picked up sequentially with a PC/PDMS stamp. The monolayer graphene is cut in two halves using an AFM tip. The two pieces are picked up with an intended rotational misalignment of $1.2^\circ$, slightly larger than the final twist angle of the device. (e) Optical image of the stack. The \wse\ crystal, tBLG, and the bottom hBN substrate are marked by red, yellow, and dark blue contours, respectively. (f)  The straight crystal edges of graphene and \WSe\ are marked with orange and red lines, respectively. 
tBLG and \WSe\ are rotationally misaligned by $44^\circ$, which is equivalent to $16^\circ$. We note that a twist angle of $16^\circ$ between tBLG and \WSe\ is expected to give rise to maximum SOC strength.
(g) Schematic of the transport measurement setup. Independent controls on charge carrier density $n_{tBLG}$ and displacement field $D$ are achieved by applying D.C. voltage to the top and bottom graphite gate electrodes, $V_{top}$ and $V_{bot}$, respectively. (h) Schematic phase diagram demonstrating the effect of the dual gated sample geometry. $D$ and $n_{tBLG}$ are obtained from $V_{top}$ and $V_{bot}$ according to Eq.~\ref{EqM3} and Eq.~\ref{EqM4}. As demonstrated in Fig.~3, SOC strength is tunable with $D$, with stronger SOC observed at positive $D$ (red shaded area). The absence of a top gate electrode confines the measurement to the dashed horizontal line along $V_{top}=0$. In this configuration, tuning the sample to electron carrier also applies a negative displacement field $D<0$, leading to weaker SOC. For example, positions of $\nu=+1$ and $+2$ are marked with black solid lines, which intersect $V_{top}=0$ in the blue shaded area indicating weak SOC. As a result, potential AHE could be undetectable in the $D<0$ regime. 
}
\end{figure*}

\begin{figure*}
\includegraphics[width=0.6\linewidth]{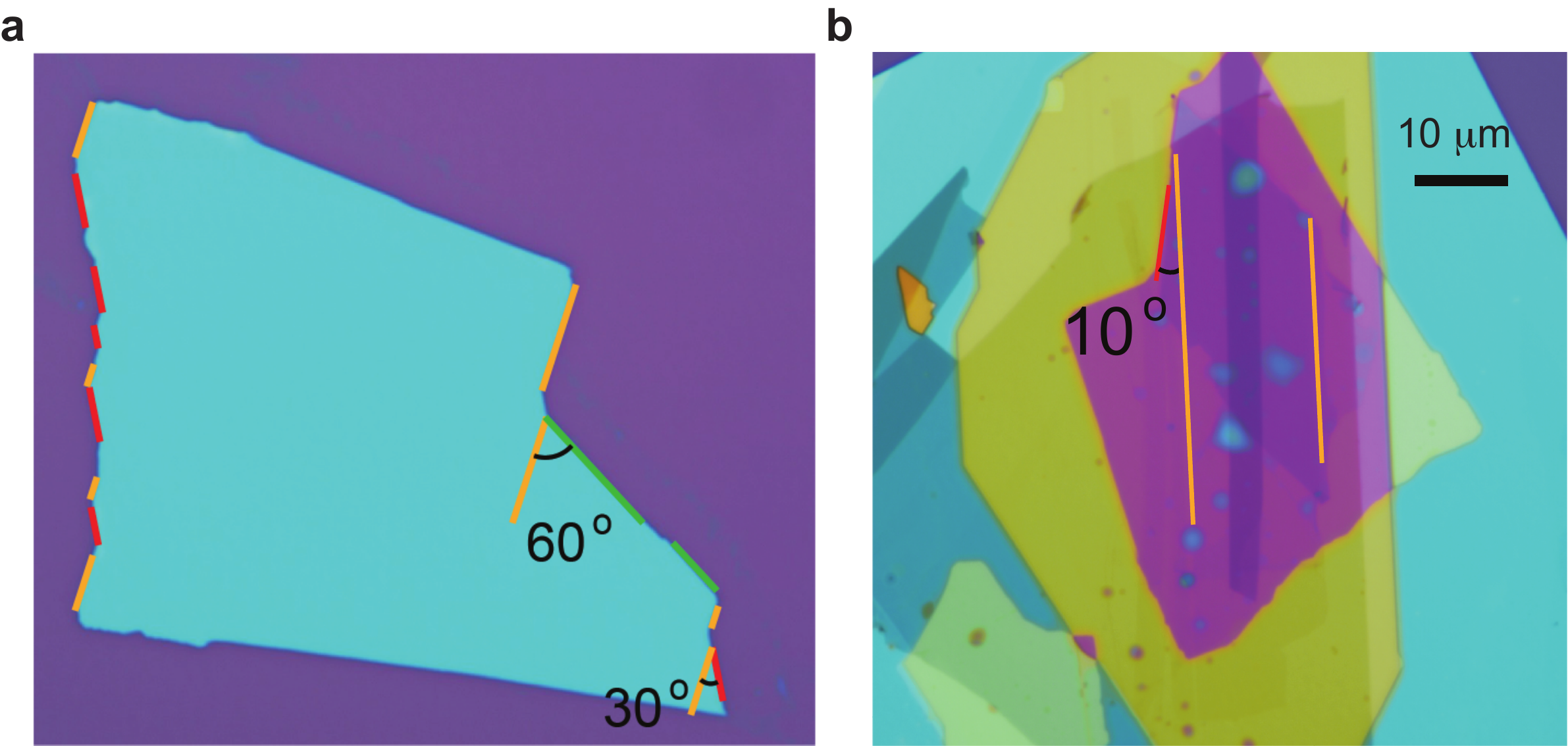}
\caption{\label{fig:crystal} {\bf{Misalignment between tBLG and hBN substrate}} (a) Optical image of the bottom hBN crystal. We identify a series of straight edges that fall into 30 degrees of misalignment with each other.  These edges likely correspond to zigzag or armchair direction of the hBN crystal axis. We mark these edges using colored lines, where lines with the same orientation are marked with the same color. The top and bottom edges of the crystal in this image do not align with any other features of the crystal. Therefore, they do not represent a true crystallographic axis of the honeycomb lattice. (b) According to the crystal axis orientation identified in (a), graphene and hBN substrates are misaligned by $10$ degrees, which is consistent with transport characterizationn in Fig.~\ref{fig:activation_gap} and Fig.~\ref{fig:compare}. 
}
\end{figure*}


All layers of two-dimensional materials used in the device are produced using the mechanical exfoliation method, which are subsequently stacked together by a poly(bisphenol A carbonate) (PC)/polydimethylsiloxane (PDMS) stamp (see Figure~\ref{FIG.fab}a-d). The tBLG is assembled using the ``cut-and-stack" technique~\cite{Saito2019decoupling}, in which a monolayer graphene is cut in half using an atomic force microscope (AFM) before being picked up to improve the twist angle accuracy and homogeneity. The layers of the device are composed of (from top to bottom): BN (36 nm), graphite (7 nm), BN (61 nm), \wse (2 nm), tBLG, BN (37 nm), graphite (5 nm). The thickness of the layers are measured with an AFM. 

The fabrication of the device follows the standard electron-beam lithography, reactive-ion etching (RIE) and electron-beam evaporation procedures. First the top graphite is removed from the contact region using CHF$_3$/O$_2$ plasma in the RIE, then the same recipe is used to define the Hall-bar shape and expose the graphene edge for the contact, finally Cr/Au (2/100nm) is deposited to form the electrodes for the tBLG and both graphite gates.

\subsection{Transport Measurement}

The dual-gated structure allows independent control of carrier density in the tBLG, $n_{tBLG}$, as well as displacement field $D$ in the out-of-plane direction. Such control is achieved by applying a DC gate voltage to top and bottom graphite electrodes, $V_{top}$ and $V_{bot}$, respectively, as shown in Fig.\ref{FIG.fab}. $n_{tBLG}$ and $D$ can be obtained using the following equations: 
\begin{eqnarray}
n_{tBLG} &=& (C_{top}V_{top}+C_{bot}V_{bot})/e+n^0_{tBLG}, \label{EqM3}\\
D&=& (C_{top}V_{top}-C_{bot}V_{bot})/2\epsilon_0, \label{EqM4}
\end{eqnarray} 
\noindent
where $C_{top}$ ($C_{bot}$) is the geometric capacitance between top (bottom) graphite and tBLG, and is determined from the conventional Hall resistance. $n^0_{tBLG}$ is the intrinsic doping in tBLG. 

Transport measurement is performed in a BlueFors LD400 dilution refrigerator with a base temperature of $15$ mK. Temperature is measured using a resistance thermometer located on the sample probe. Standard low frequency lock-in techniques with Stanford Research SR830 amplifier are used to measure resistance $R_{xx}$ and $R_{xy}$, with an excitation current of $2$ nA at a frequency of $17.77$ Hz. An external multi-stage low-pass filter is installed on the mixing chamber of the dilution unit. The filter contains two filter banks, one with RC circuits and one with LC circuits. The radio frequency low-pass filter bank (RF) attenuates above $80$ MHz, whereas the low frequency low-pass filter bank (RC) attenuates from $50$ kHz. The filter is commercially available from QDevil.

The parallel- and tilted-field measurement is performed by mounting the device on a homemade adapter which fixes the device at a desired angle relative to the field orientation. The tilting angle is further determined by extracting the perpendicular component of the magnetic field from the conventional Hall resistance (Fig.\ref{fig:Bperp}).

The atomically-thin \wse\ has a large band gap and thus becomes insulating at cryogenic temperature. Therefore, it can be viewed as part of the dielectric layer, and does not contribute to the electrical transport signal directly.~\cite{island2019spin}

\begin{figure*}
\includegraphics[width=1\linewidth]{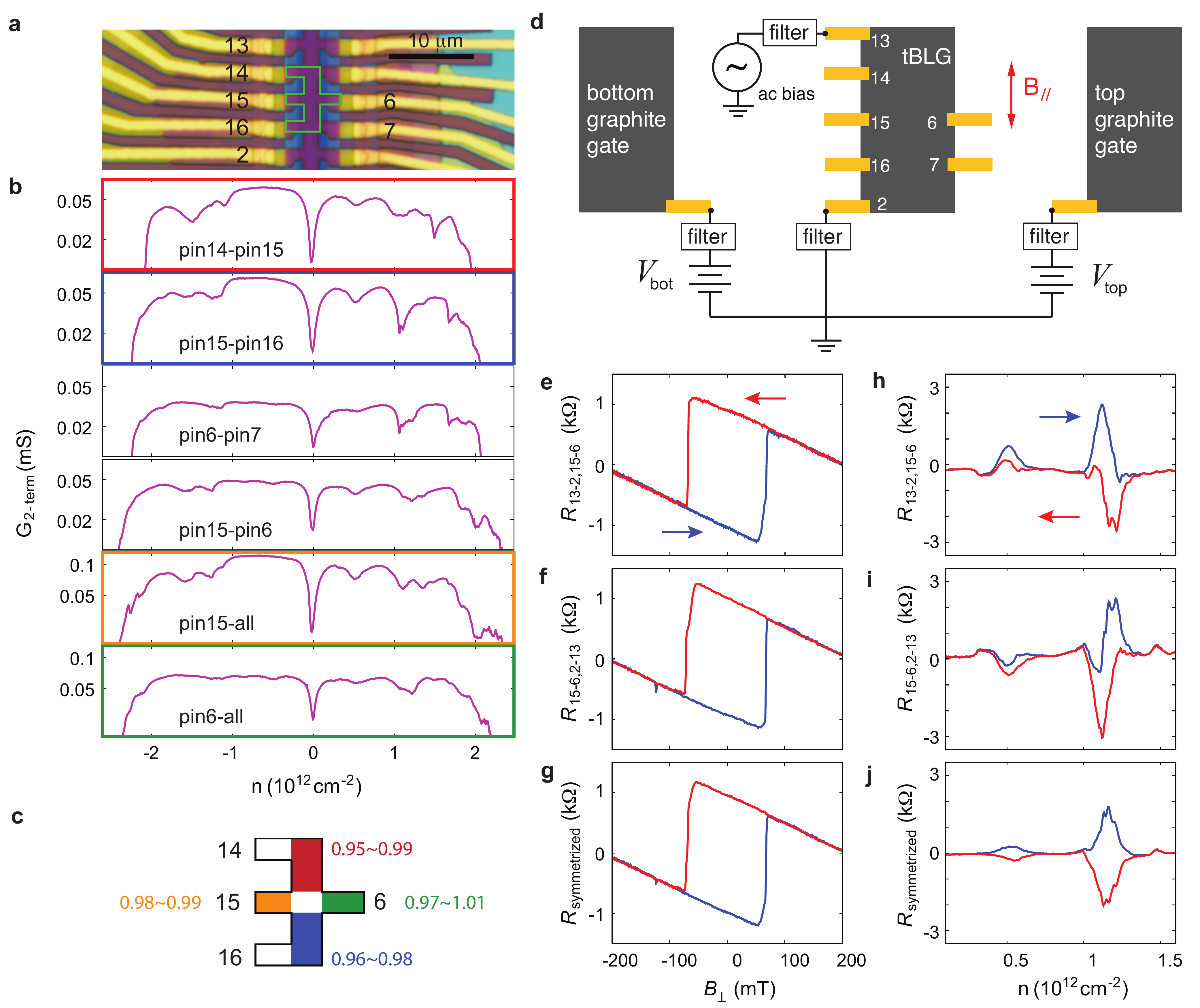}
\caption{\label{fig:2term} {\bf{Transport measurement }} (a) Optical image of the device. Pins used in the measurement are labeled with numbers. The green contour marks the region where the twist angle is near the magic angle. All data presented are taken within this region. Pin 13 and 2 are used as the current injection, pin 15 and 16 are used as longitudinal voltage probes, and pin 15 and 6 are used as Hall voltage probes. (b) Two-terminal conductance $G_{2-term}$ is measured with fixed bottom gate $V_b=0$ at 20mK and $B_\perp=0.4T$. Vertical axes are in log scale. (c) Twist angles are labeled for different parts of the outlined area in (a). (Unit: degree) The colors are corresponding to the panels in (b). Each twist angle is determined from the density difference between CNP and the $\nu=+2$ CI state, and the broadening of the conductance minimum accounts for the uncertainty. According to these values, we attribute the twist angle for the area of interest to be $0.98^\circ\pm0.03^\circ$. 
(d) Schematic of transport measurement. Red arrow marks the direction of the in-plane magnetic field. Low-pass filtering is used for measuring resistance as well as applying DC voltage bias. Charge carrier density $n_{tBLG}$ and displacement field $D$ are controlled by applying DC voltage bias to top graphite and bottom graphite. In the following we detail the scheme of performing geometric symmetrization for measuring hysteresis loops.
(e-g) Symmetrization of magnetic field-induced hysteresis loop at $\nu=+2$, measured at $T=2$~K and $D=0$. In (e), current is injected from pin 13 to pin 2, and Hall voltage is measured from pin 15 to pin 6. We label this configuration as $R_{13-2,15-6}$. In (f), current is injected from pin 15 to pin 6, and Hall voltage is measured from pin 2 to pin 13. (g) Shows the symmetrized Hall resistance using Onsager reciprocal relation as $R_{symmetrized}=(R_{13-2,15-6}+R_{15-6,2-13})/2$. (h-j) Symmetrization of density-induced hysteresis loop at both $\nu=+1$ and $\nu=+2$, measured at $T=20$~mK, $D=252$~mV/nm and $B_{\perp}=10$~mT.
}
\end{figure*}

\begin{figure*}
\includegraphics[width=0.68\linewidth]{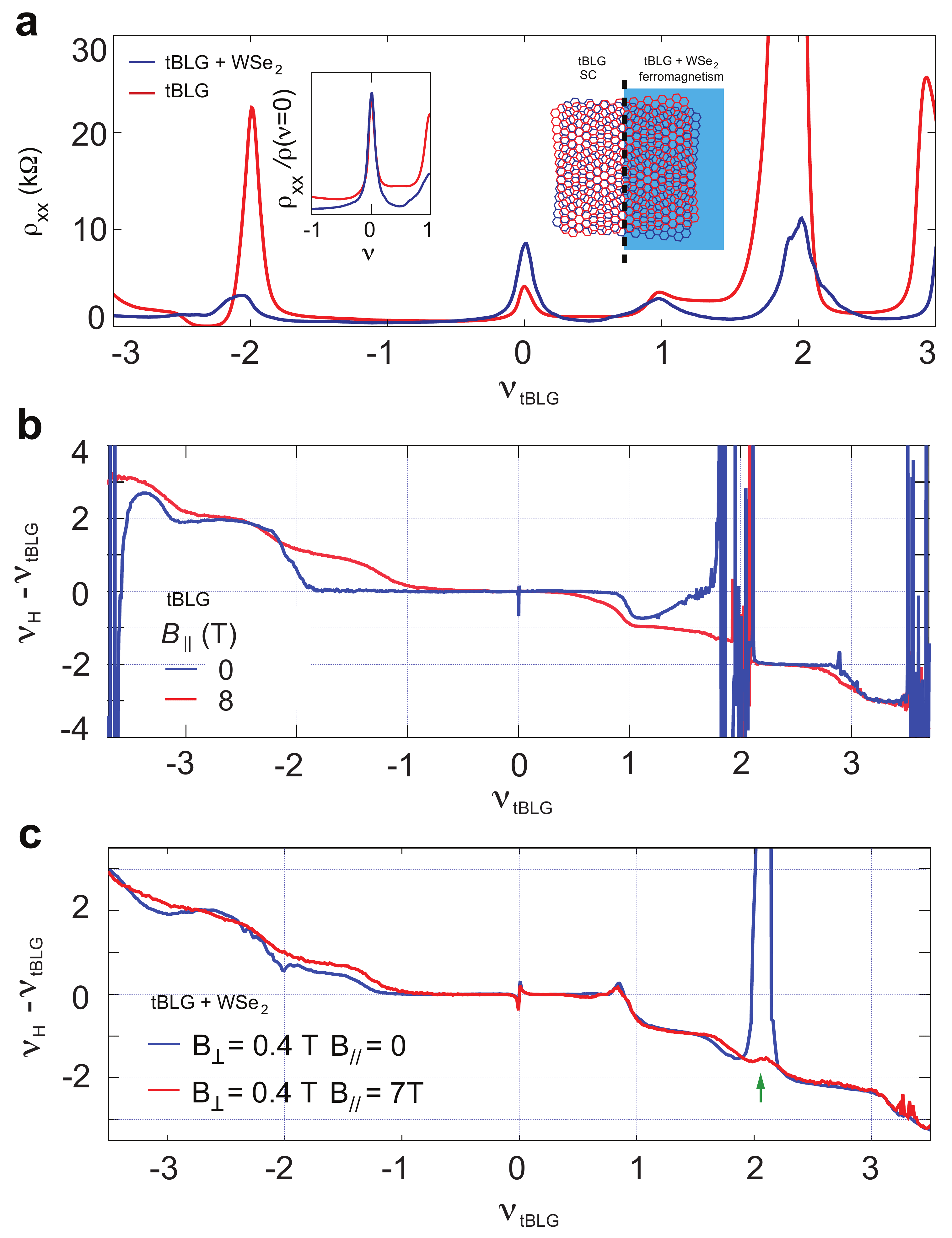}
\caption{\label{fig:compare} {\bf{Comparison between tBLG devices with and without \wse}} (a) Longitudinal resistivity $\rho_{xx}$ as a function of moir\'e filling, measured from two samples at $T=20$ mK and $B = 0$. The sample without tBLG/\WSe\ interface (red trace) exhibits a robust superconducting phase on the hole doping side of $\nu=-2$, whereas superconductivity is absent in the sample with tBLG/\WSe\ interface (blue trace). Left inset, $\rho_{xx}$ renormalized by the peak value at the charge netrality point. Notably, these two samples show the same broadening around the charge neutrality point, indicating comparable disorder concentration and similar sample quality. As such, the absence of the superconducting phase is not caused by difference in sample quality. Right inset: schematic sample geometry for creating a SC/ferromagnetism boundary along the atomic edge of the WSe2 crystal, by partially covering tBLG with a WSe2 crystal, ferromagnetism is stabilized by the tBLG/WSe2 interface on the right side of the sample, whereas SC would emerge on the left in the absence of SOC. 
(b) Normalized Hall density measured from magic angle tBLG with $\theta = 1.06^{o}$ without SOC. The effect of a large in-plane $B$-field introduces an extra step at $\nu=-1$, denoting symmetry breaking in iso-spin polarization. (c) Normalized Hall density measured from magic angle tBLG with $\theta = 0.98^{o}$ with proximity-induced SOC. The step at $\nu=-1$ is present at $B=0$, which remains mostly the same in the presence of large in-plane $B$. Comparing with panel (b), Hall density measurements indicate that the effect of SOC on isospin polarization is similar to that of a large in-plane $B$-field. The divergence in Hall density measured at \Bpara $=0$, marked by the green arrow, is due to the anomalous Hall effect, which is suppressed by a large in-plane $B$.
}
\end{figure*}

\begin{figure*}
\includegraphics[width=0.8\linewidth]{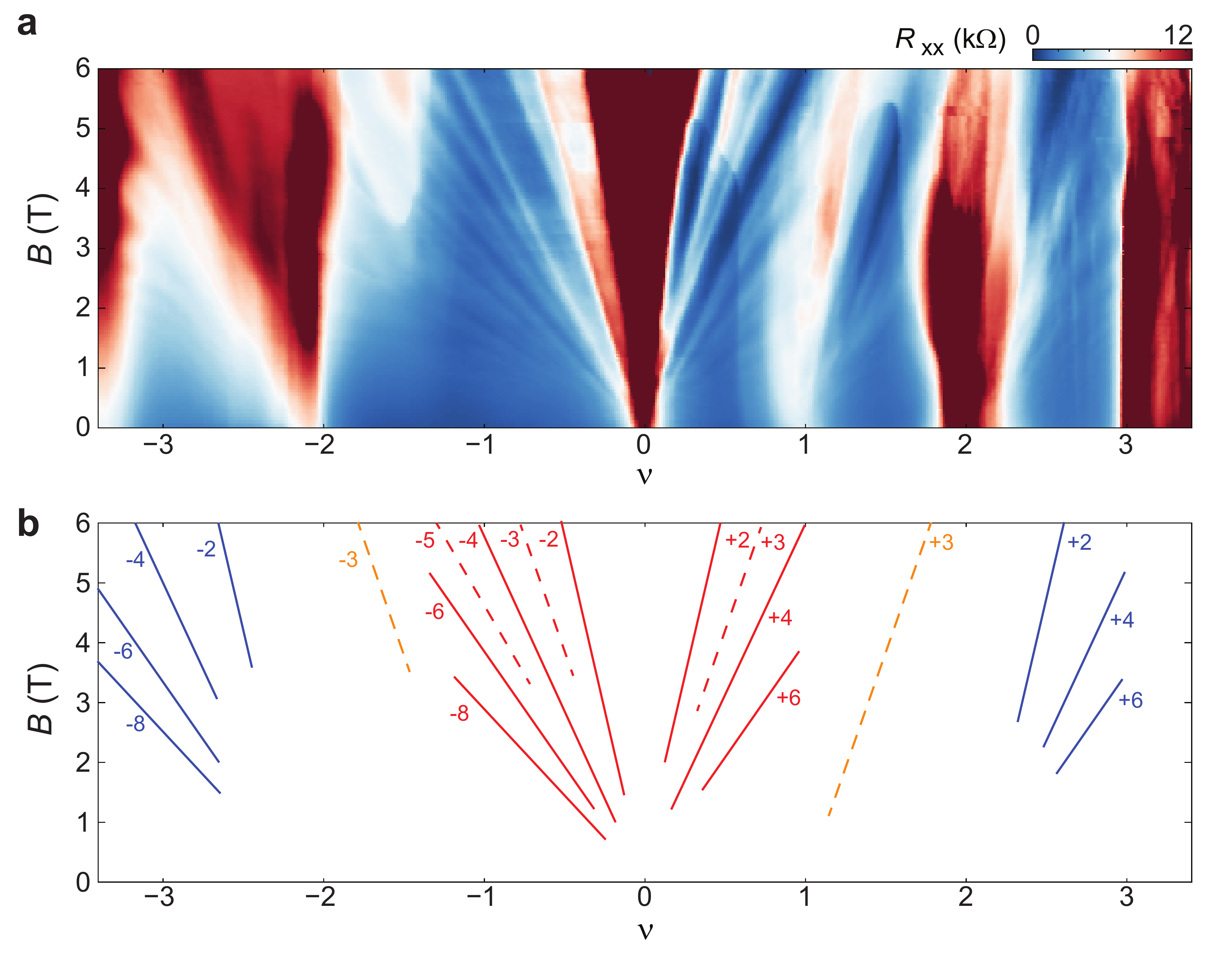}
\caption{\label{fig:fan} {\bf{Landau fan}} (a) \Rxx\ measured as a function of out-of-plane magnetic field \Bperp\ and moir\'e filling, measured at $D=252$~mV/nm, $B_{//}=0$ and $T=25$~mK. (b) Schematic showing the most prominent features in Landau fans emerging from integer moir\'e fillings and the CNP. Landau levels originating from CNP, $\nu= \pm$1, and $\pm$2 are labeled with red, orange, and blue colored lines, respectively. Landau levels with even (odd) Chern number are labeled with solid (dashed) lines. The hierarchy and degeneracy of Landau fans emerging from integer moir\'e fillings are similar to magic-angle tBLG without alignment to hBN substrate ~\cite{Cao2018b,Yankowitz2019SC,Saito2020pomeranchuk}, providing further confirmation that graphene/hBN coupling does not play a role in the observed orbital ferromagnetism in our sample.}
\end{figure*}

\begin{figure*}
\includegraphics[width=0.7\linewidth]{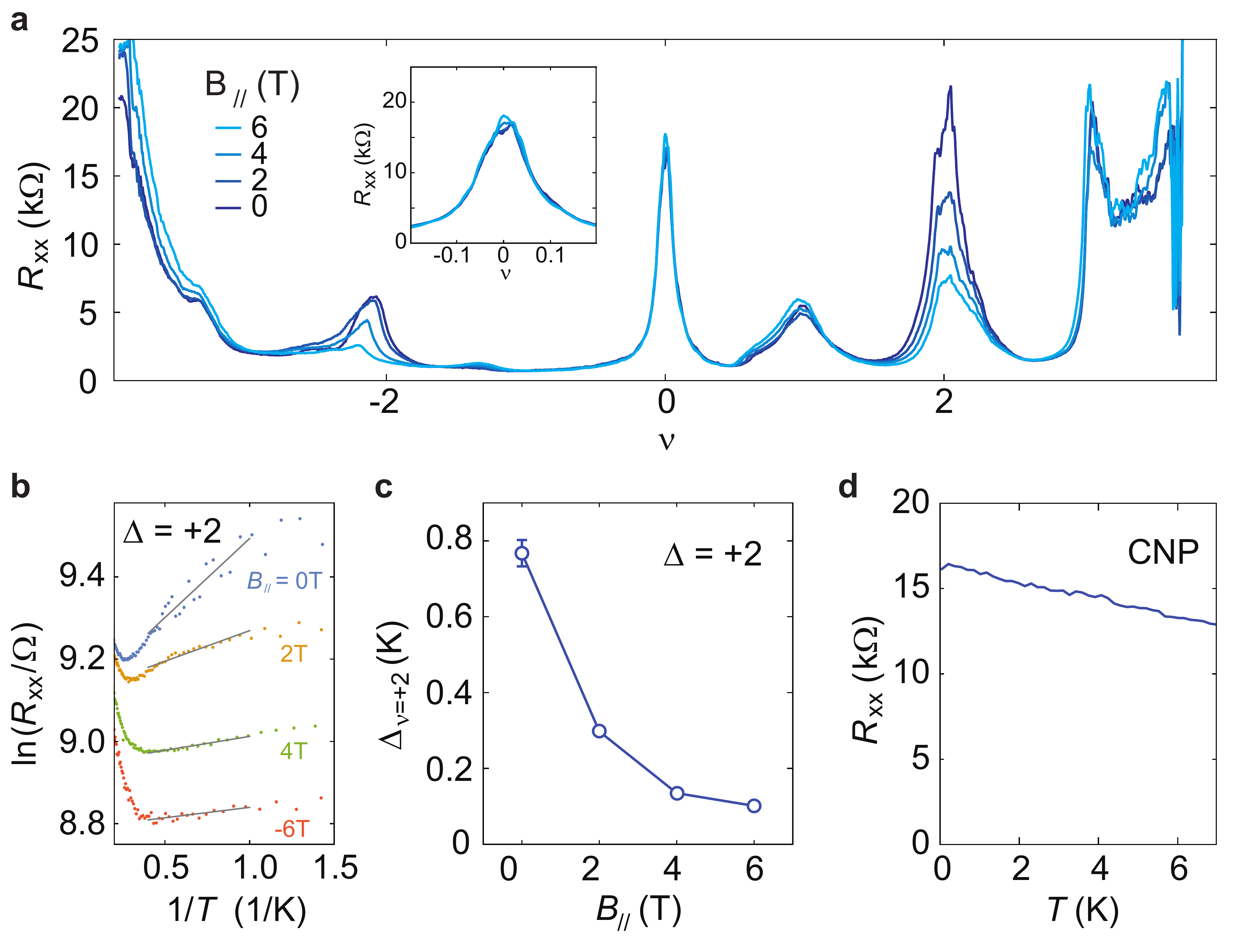}
\caption{\label{fig:activation_gap} {\bf{Parallel-field and temperature dependence of the insulating states}} (a) Longitudinal resistance \Rxx\ versus moir\'e band filling factor $\nu$ measured at different parallel magnetic fields $B_{\parallel}$ at 20mK. In this measurement, we fix the applied voltage on the bottom gate at $0$ V, and the carrier density is tuned by sweeping the applied voltage on the top gate. The CI state at $\nu=2$ becomes weaker with increasing \Bpara, indicative of a spin-unpolarized ground state. The inset shows the  peak in \Rxx\ at the CNP, which remains the unchanged in the presence of an in-plane $B$-field up to $6$ T. This rules out a sublattice gap near the CNP, indicating that graphene and hBN substrate are not aligned ~\cite{Zibrov2018even}.  (b) Arrhenius plot of \Rxx\ at $\nu=+2$ at different $B_{\parallel}$. The measurements at $B = 0$, $2$ and $4$ T are measured with $D=252$ mV/nm, whereas the measurement  at $-6$ T with $D=113$ mV/nm. (c) The energy gap of the CI at $\nu=2$, $\Delta_{\nu=+2}$, which is  extracted from the activated behavior shown in (b), shows a monotonic decline  with increasing \Bpara. (d) \Rxx\ versus temperature measured at the CNP down to the base temperature $T = 20$ mK. The absence of thermally activated behavior indicates the Dirac point for tBLG is gapless, providing direct evidence that tBLG is not aligned with the hBN substrate. 
}
\end{figure*}


\begin{figure*}
\includegraphics[width=0.68\linewidth]{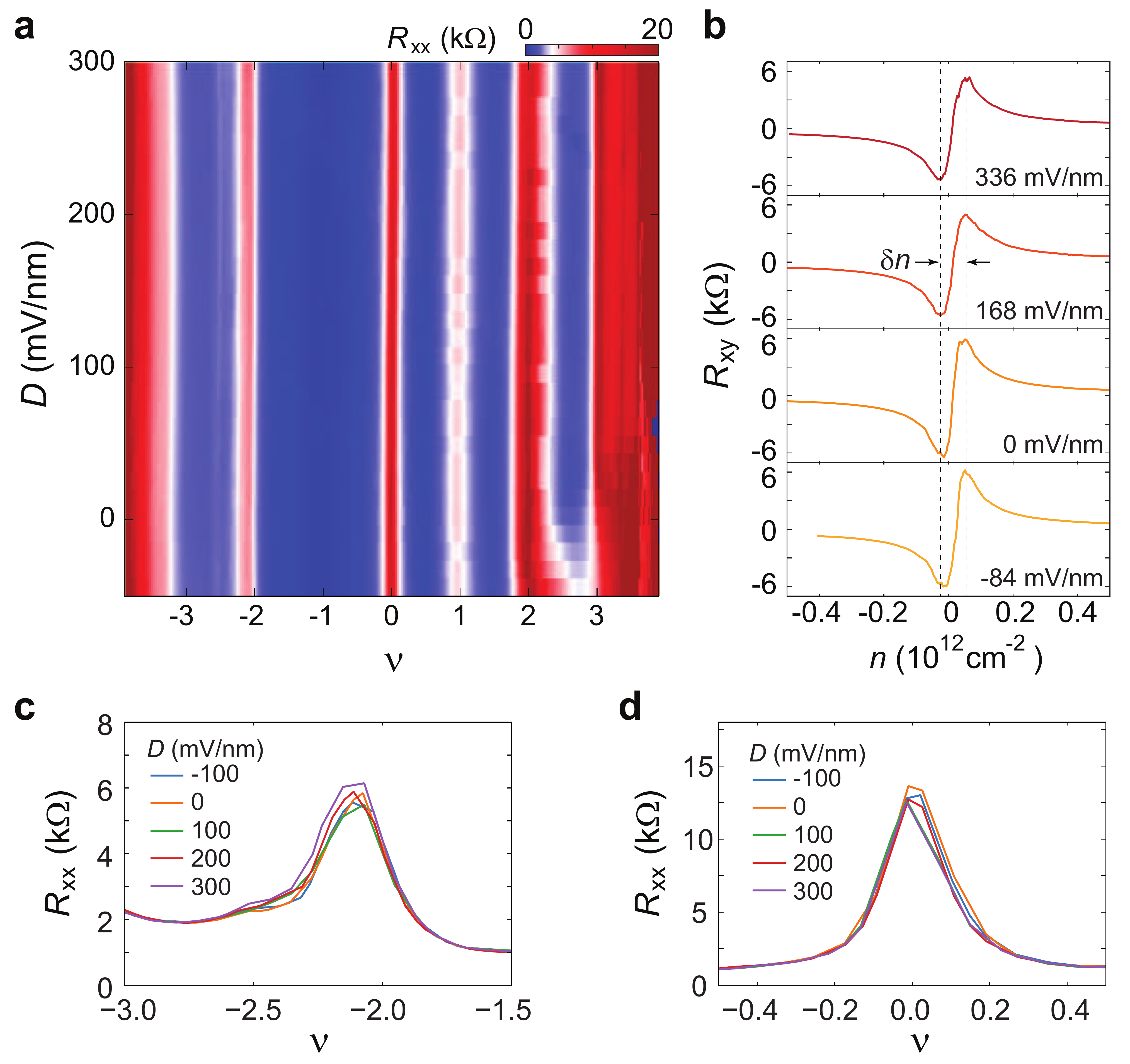}
\caption{\label{fig:Rxy_at_CNP} {\bf{Displacement field dependence}} (a) Longitudinal resistance \Rxx\ as a function of displacement field $D$ and moir\'e filling factor $\nu$. The measurement is performed at 20mK and $B=0$. The value of $D$ and $\nu$ are determined according to Eq.~\ref{EqM3} and Eq.~\ref{EqM4}. (b) Hall resistance \Rxy\ versus carrier density $n$ measured at different displacement fields $D$ near the charge neutrality point. Traces shown are anti-symmetrized with data taken at $\pm$400mT as \Rxy$=(R_{xy, +400mT}-R_{xy, -400mT})/2$. The width of the disordered regime, defined as the density difference between the extrema in \Rxy, $\delta n$, is around 0.8 (10$^{11}$cm$^{-2}$), indicating excellent sample quality. The fact that $\delta n$ remains the same over a wide range of $D$ suggests that sample quality does not depend on layer polarization and two graphene layers exhibits comparable levels of disorder. As such, we conclude that the effect of $D$ on the AHE does not originate from the layer dependence of disorder or strain. (c-d) Linecuts of \Rxx\ as a function of $\nu$ measured at different $D$ near $\nu=-2$ (c) and CNP (d). The superconducting phase is absent over the full $D$ range. Data taken at 20mK and $B=0$.
}
\end{figure*}

\begin{figure*}
\includegraphics[width=0.9\linewidth]{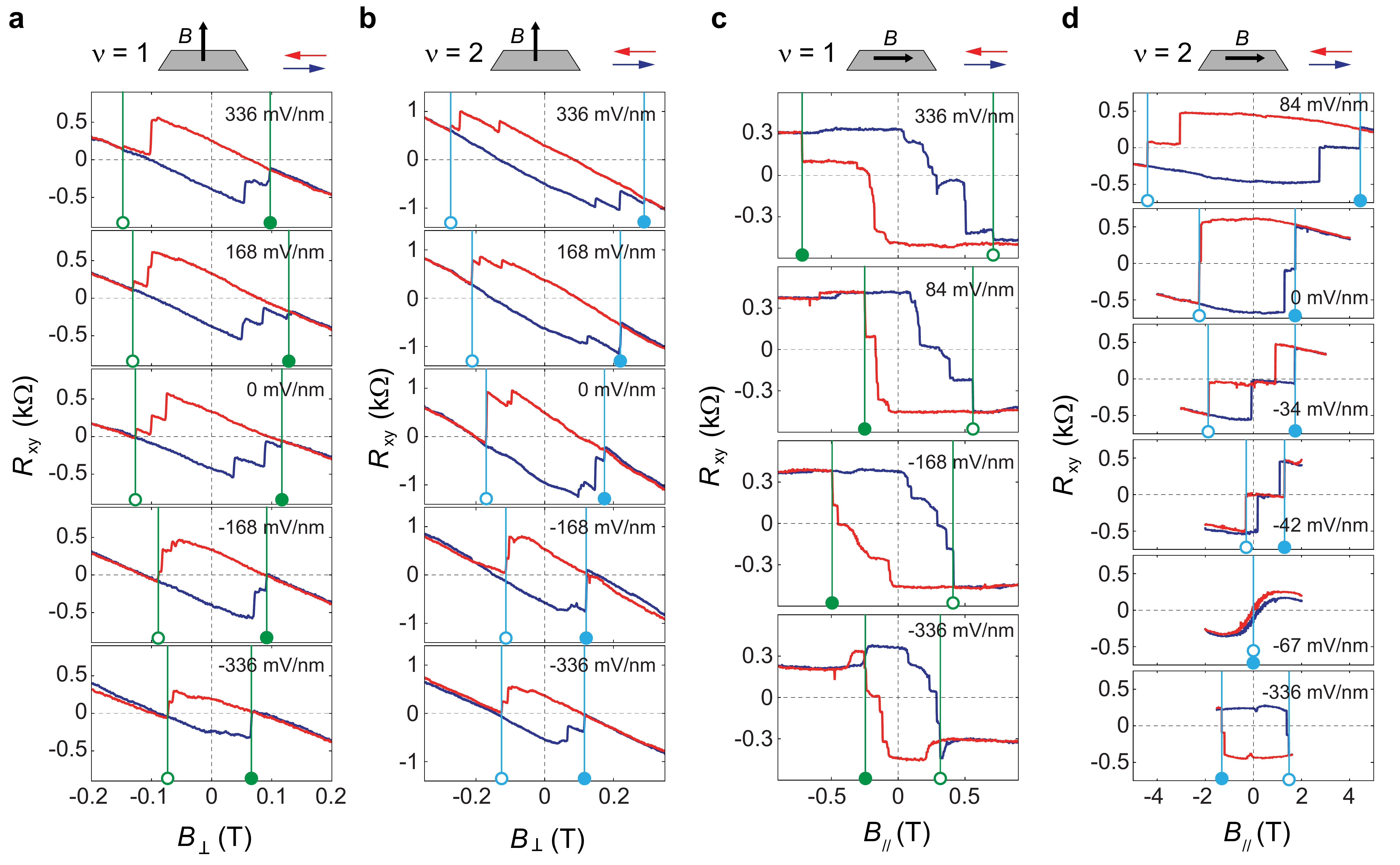}
\caption{\label{fig:Bloop} {\bf{$B$-induced hysteresis loop in \Rxy}} Hall resistance as a function of (a-b) out-of-plane and (c-d) in-plane magnetic field. The measurement is performed at moir\'e filling of (a) and (c) $\nu=1$, (b) and (d) $\nu=2$. Vertical lines with solid circle represent \Bupperp\ and \Bupp, whereas vertical lines with open circle denote \Bdownperp\ and \Bdownp. (a-c) are measured at $T= 20$ mK. (d) is measured at $T=3$~K with the exception of the bottom panel, which is taken at $T=1.5$~K.}
\end{figure*}

\begin{figure*}
\includegraphics[width=0.9\linewidth]{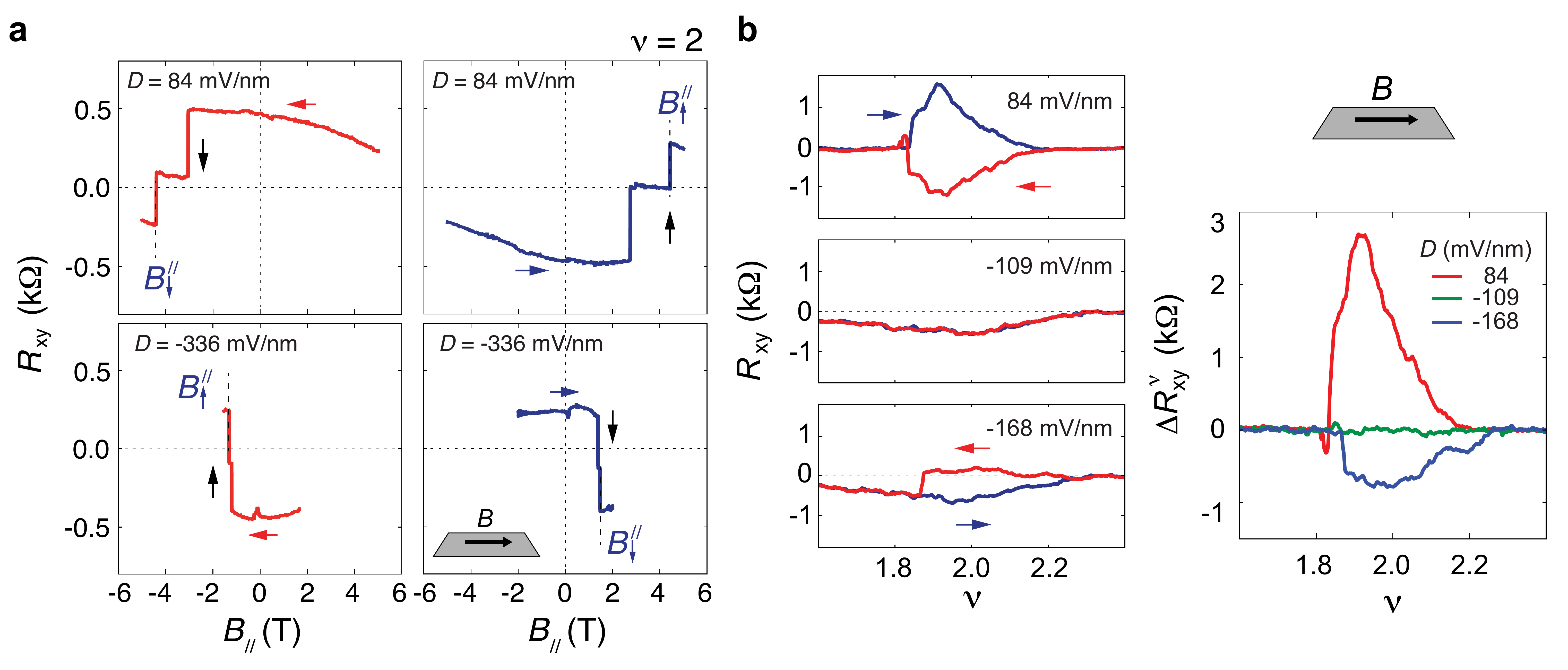}
\caption{\label{fig:D} {\bf{$D$-induced magnetization reversal}} (a) Varying $D$ induces a magnetization reversal in the presence of an in-plane $B$-field, which is demonstrated in (a) magnetic field and (b) doping induced hysteresis loops. (a) In $B$-induced hysteresis loops, we define coersive field \Bup\ (\Bdown) as the value of $B$ where the sign of \Rxy\ switches from negative to positive (positive to negative). Given a fixed direction of sweeping $B$, \emph{e.g.}, from positive to negative (left two panels),  Hall resistance at $B=0$ changes sign at different $D$. Owing to the specific definition of the coercive field, sign change in Hall resistance gives rise to sign change of \Bup\ and \Bdown\ accordingly. (b) In the presence of an in-plane field, doping-induced hysteresis loops exhibit the same $D$ dependence as $B$-induced switching. A sign change in \DRxyn\ is observed at $D\sim -100$ mV/nm.}
\end{figure*}

\begin{figure*}
\includegraphics[width=0.8\linewidth]{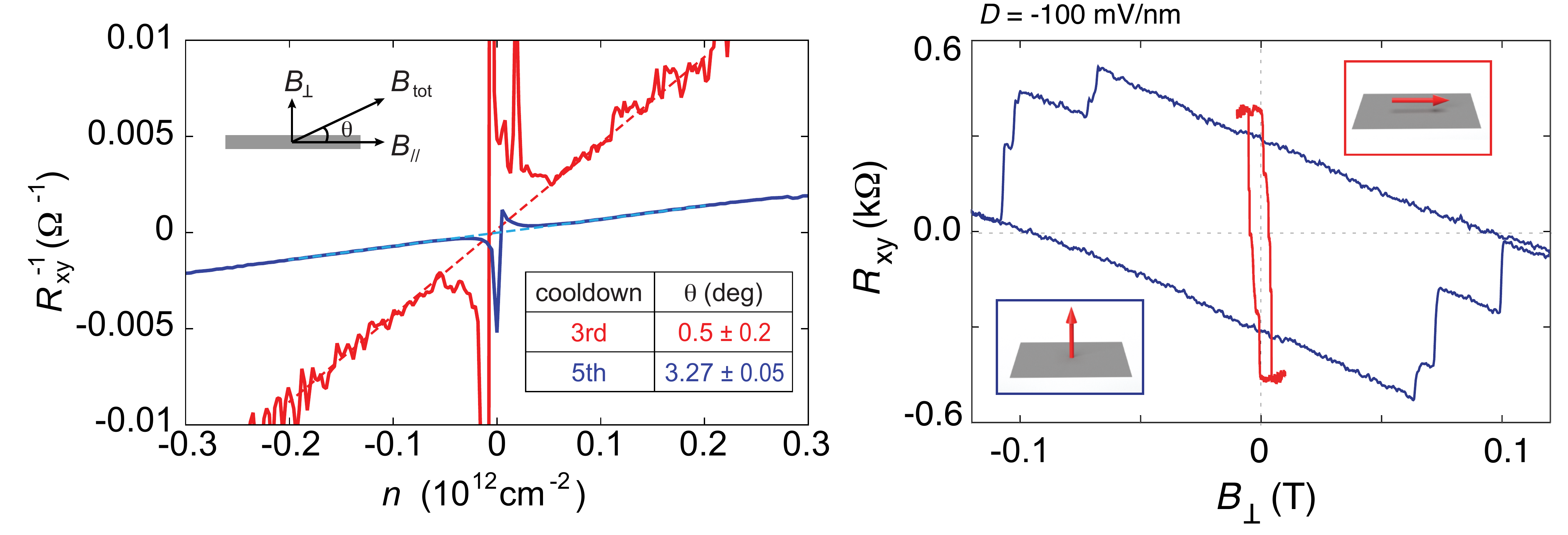}
\caption{\label{fig:Bperp} {\bf{Comparing in-plane and out-of-plane $B$}} Left: inverse Hall resistance as a function of  carrier density $n$, measured with an external magnetic field of $B_{tot}=4$~T. Since Hall resistance is determined by carrier density $n_{tBLG}$ and the out-of-plane component of $B$, $1/R_{xy}=ne/B_\perp$, the linear slope of $1/R_{xy}$ provides an accurate measurement for \Bperp. Based on this, we calculate the tilt angle between $B$ and the 2D plane using $\theta=$sin$^{-1}(B_\perp/B_{tot})$. The measured tilting angle is shown in the chart. 
All in-plane B-field measurements in the main text are performed during the 3rd cool down (solid red curve). We intentionally misaligned the $B$-field to have a small out-of-plane component in the 5th cool down (solid blue curve).  
Right: Hall resistance as a function of the out-of-plane component of $B$. Blue and red traces denote Hall resistance measured with $B$ aligned $90^{o}$ (perpendicular) and $0.5^{o}$ (predominantly in-plane) from the 2D plane. The measurements are performed at $\nu=+1$, $T=20$ mK and $D = 100$ mV/nm. For the in-plane measurement, the value of out-of-plane $B$ component at the magnetization reversal is orders of magnitude smaller than the out-of-plane coercive field, suggesting the influence of the out-of-plane $B$ component is negligible. Instead, magnetization couples to the external $B$ field through the combination of spin Zeeman and SOC. 
 }
\end{figure*}

\begin{figure*}
\includegraphics[width=0.65\linewidth]{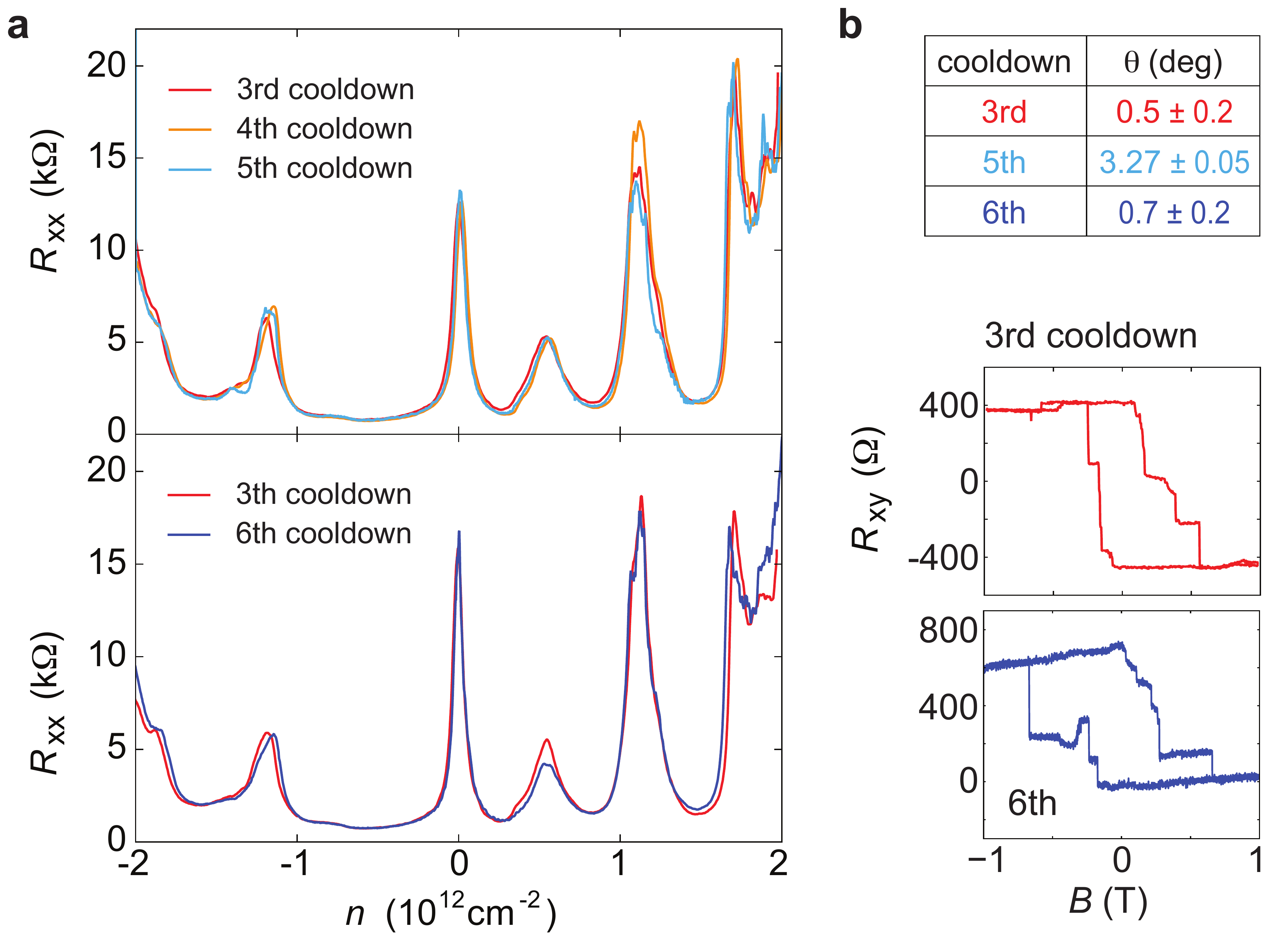}
\caption{\label{fig:cooldown} {\bf{Compare between different cooldowns.}} (a) Longitudinal resistance \Rxx\ as a function of carrier density, measured at four consecutive cooldowns, labeled as 3rd, 4th, 5th, and 6th cooldowns. The device is placed in a parallel, perpendicular, tilted, and parallel orientation with regard to the applied magnetic field in the 3rd, 4th, 5th, and 6th cooldown, respectively. Data shown in the upper panel in (a) is taken at $D=252$~mV/nm, $T=0$, and $B=0$, and the data in the lower panel is taken with fixed $Vb=0$ at $T=1$~K, and $B=0$. The twist angle of the device is stable between thermal cycles, as the resistance peaks at the CI states overlap for these curves. (b) 
Top: the tilting angle $\theta$ between the external magnetic field and the device plane for different cool down, which is obtained by measuring Hall resistance as a function of carrier density (Fig.~\ref{fig:Bperp}). Bottom: \Bpara-induced hysteresis loops measured at $\nu=+1$ from the 3rd and the 6th cooldown. Here, hysteresis loop from the 3rd cooldown is measured at $D=84$~mV/nm and $n=0.54$\densityunit\, whereas the 6th cooldown is measured at $D=56$~mV/nm and $n=0.49$\densityunit. Two hysteresis loops show similar behaviors. Subtle differences likely arise from the lack of geometrical symmetrization in the 6th cooldown.  Note that all \Bpara\ measurements in the main text are performed in the 3rd cooldown. }
\end{figure*}

\subsection{Doping and current-induced hysteresis}


In the presence of an in-plane $B$-field, magnetization reversal can also be controlled by sweeping the field effect tuned density. 
Doping-induced hysteresis loops are evidenced by non-zero residual Hall resistance, \DRxyn, which is observed in the $B$-field range between the coercive fields (Fig.~\ref{fig:gatecurrentloop}b-c). Since the ground state at $\nu=2$ is an orbital magnet, doping control on the magnetic order is consistent with a recent theoretical proposal for orbital magnetism in a moir\'e lattice ~\cite{Zhu2020AHE}. In the same vein, doping-induced magnetization reversal at quarter filling of the moir\'e band (Fig.~1c) suggests a similar orbital origin for the magnetic order.  In addition, magnetization reversal can be induced by changing D.C. current bias in the presence of an in-plane magnetic field.

\begin{figure*}
\includegraphics[width=1\linewidth]{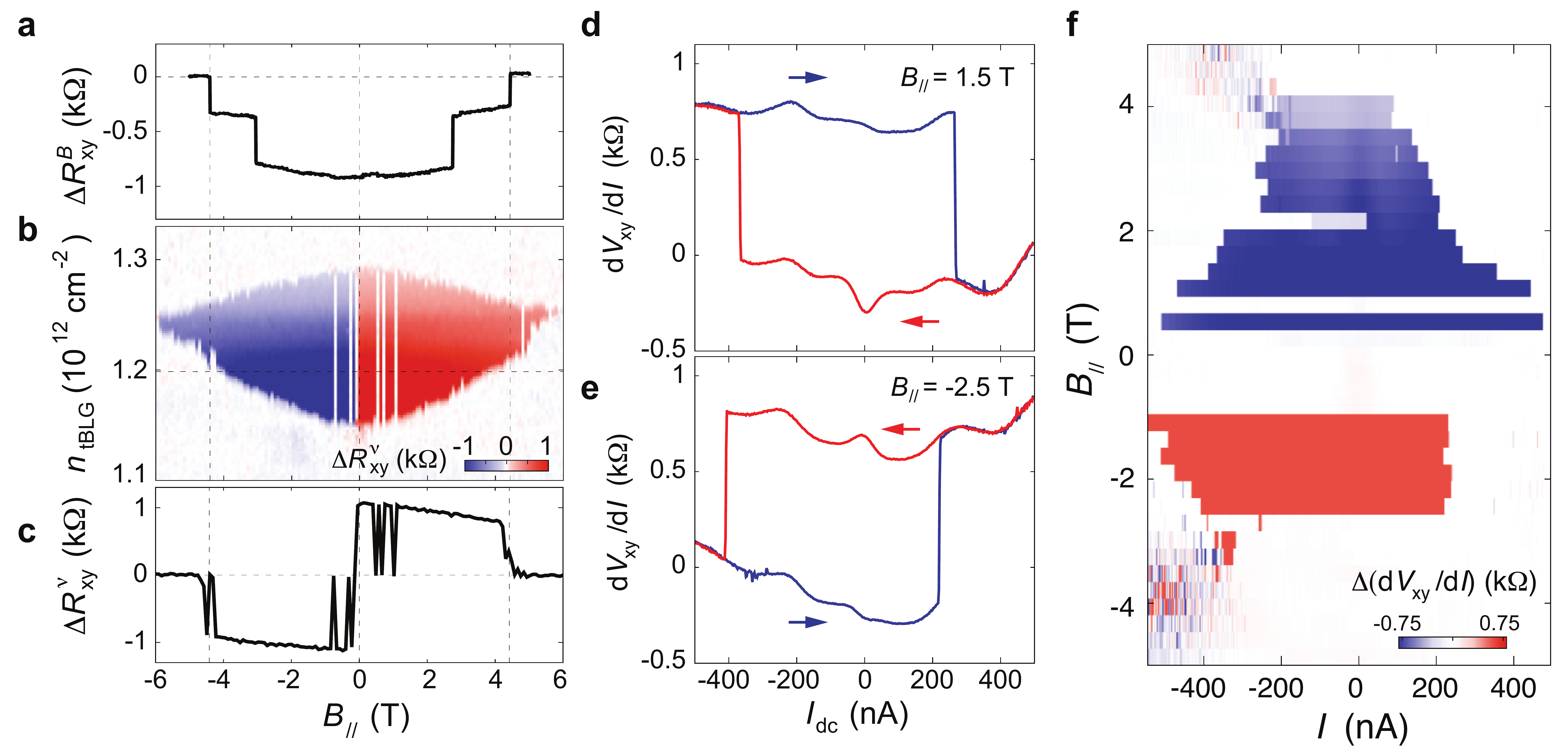}
\caption{\label{fig:gatecurrentloop} {\bf{Magnetization reversal in doping and current-induced hysteresis loops.}} (a-c) Residual Hall resistance  \DRxyb (\DRxyn), defined as the difference of Hall resistance when $B$ (density) is swept up and down. (a) \DRxyb\ versus \Bpara. Vertical dashed lines mark the coercive field from the $B$-induced hysteretic switching. (b) \DRxyn\ as a function of \Bpara\ and $n_{tBLG}$. (c) Line trace taken from panel (b) at $n_{tBLG}= 1.2$\densityunit (marked by the honrizontal dashed line). Results in (a-c) are measured at $T=3$~K and $D=85$~mV/nm. (d-e) Current-induced hysteresis loop at \Bpara$=1.5$~T (d) and \Bpara$=-2.5$~T (e). $dV_{xy}/dI$ is the differential Hall resistance measured at $T=30$~mK, $D=-85$~mV/nm and $n_{tBLG}=1.27$\densityunit. (f) Residual differential Hall resistance as a function of \Bpara\ and d.c. bias current, showing a sign change around \Bpara $ = 0$. This observation indicates that the magnetic order is determined by the combination of d.c. current and the external magnetic field. Note that current-induced hysteresis loop vanishes near \Bpara $ = 0$, which may be attributed to domain formation since magnetization reversal is expected to occur in this regime. }
\end{figure*}

\begin{figure*}
\includegraphics[width=1\linewidth]{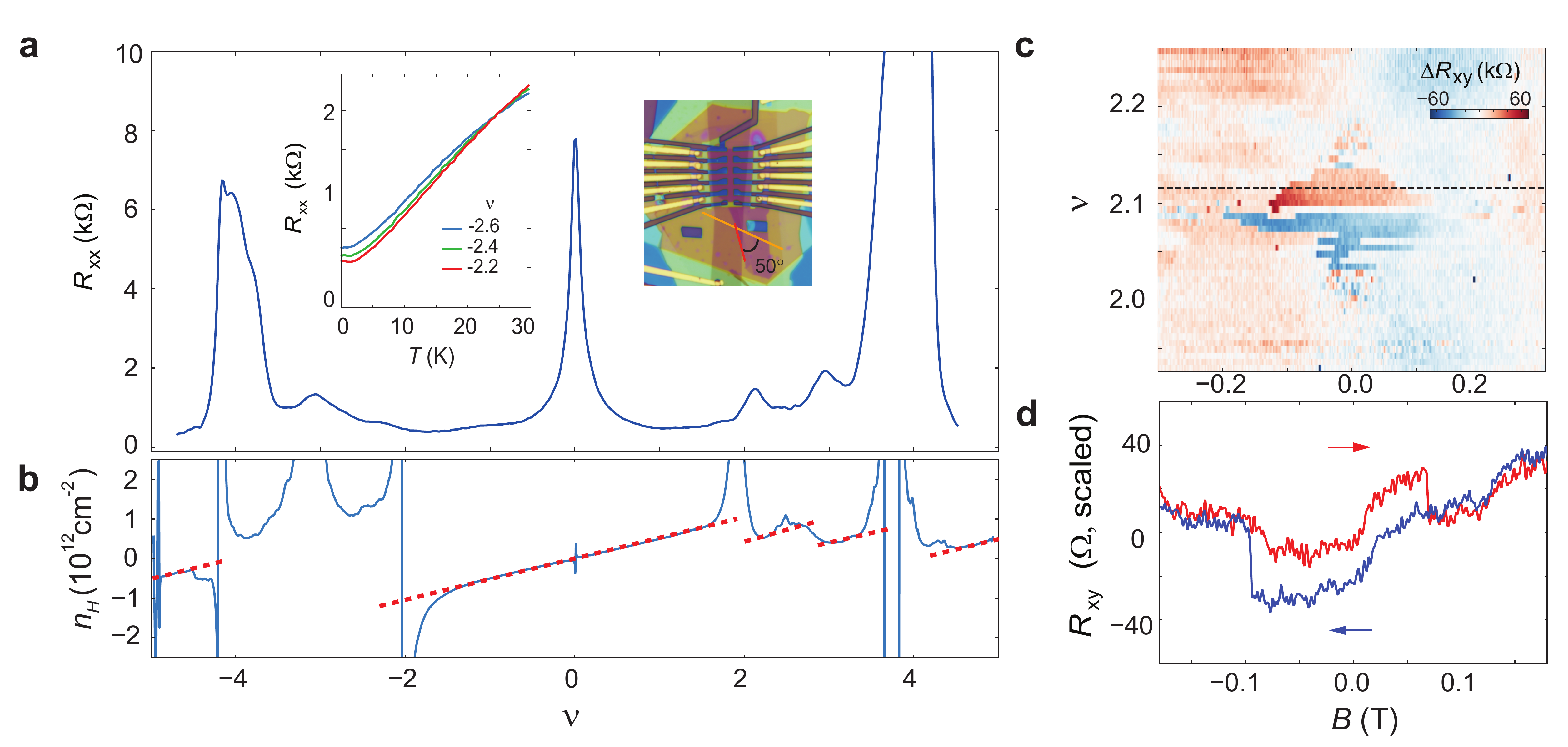}
\caption{\label{fig:erin} {\bf{Additional device A1}} (a-b) Longitudinal resistance \Rxx\ and Hall density $n_H$ measured from sample A1 as a function of moir\'e filling factor. \Rxx\ and Hall density are measured at $B=0$ and $0.8$ T, respectively. \Rxx\ shows weakly developed resistance peaks at integer fillings, which coincide with either resets or van Hove singularities in the Hall density. Right inset, optical image of sample A1, which has the same structure (\WSe\ on top of tBLG) as shown in Fig.~1a. The twist angle for tBLG is $0.93^{\circ}$, whereas \WSe\ (straight crystallographic edge marked by red line in figure) and graphene (orange) are misaligned at $10^\circ \pm 5^\circ$, which is expected to give rise to strong SOC. Left inset, $R_{xx}$ as a function of temperature measured at different moir\'e fillings on the hole doping side of $\nu=-2$. (c)  Residual Hall resistance \DRxy\ shows hysteretic switching behavior around $\nu=+2$. Note that the hysteresis loops changes sign with varying density, which is reflected in a sign change in \DRxy. This is consistent with the expected behavior of an orbital Chern insulator, which is previously observed in mono+bilayer graphene moir\'e structures ~\cite{Polshyn20201N2,Zhu2020AHE}. (d) Linecut from panel (c) shows the magnetic hysteresis loop. It is worth pointing out that the CI at $\nu=+2$ is not well developed, indicating that sample bulk is conductive, which contributes to the small size of the hysteresis loop in (d). 
}
\end{figure*}

\begin{figure*}
\includegraphics[width=0.75\linewidth]{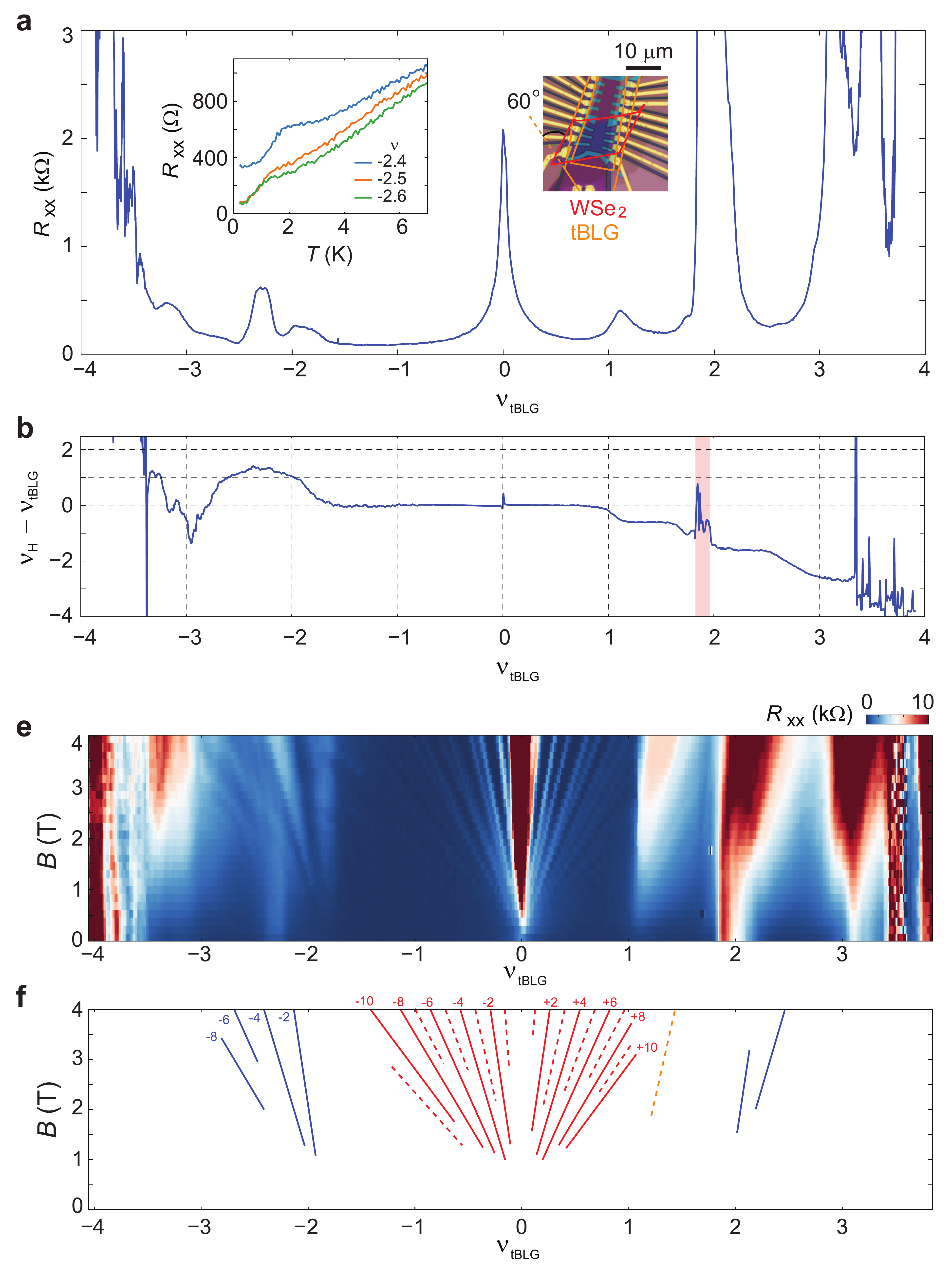}
\caption{\label{fig:zhi} {\bf{Additional device A2}} (a)  Longitudinal resistance \Rxx\ as a function of moir\'e filling $\nu_{tBLG}$, measured at $T=25$~mK and $B=0$ for sample A2. Correlated insulators are observed at $\nu=-2$, $+1$ and $+2$, which is similar to the sample reported in the main text. The twist angle of tBLG is calculated based on the position of CIs to be $1.08^{\circ}$. Left inset: \Rxx\ as a function of temperature measured at $B=0$ and different fillings. Although \Rxx\ shows a downturn at low temperature, a zero resistance state is not observed even at the lowest temperature of $T=20$ mK. This is in stark contrast with the observed behavior for tBLG without SOC at this twist angle, where robust superconducting phase is expected to have transition temperature of $T_c \sim 2$ K). Right inset: optical image of sample A2 with the same stacking geometry as shown in Fig.~1a (\WSe\ on top of tBLG).  The orange and red contours outline the WSe$_2$ and tBLG flakes, showing perfect alignment between \WSe\ and graphene, which is expected to yield weaker SOC strength ~\cite{Koshino2019SOC}.  (b) Renormalized Hall density as a function of moir\'e filling, measured at $T=25$~mK and $B=0.15$~T. The red shaded stripe marks a feature in the Hall density that corresponds to the onset of anomalous Hall signature (see Fig.\ref{fig:compare} for comparison). (c) \Rxx\ as a function of perpendicular magnetic field and moir\'e filling. Landau fans emerging from integer fillings show similar behavior compared to tBLG near the magic angle without SOC.  (d) Schematic of most prominent features of Landau fans shown in panel (c). Red, orange, and blue lines denote Landau fans originating from CNP, $\nu=+1$, and $\nu=\pm2$, respectively. Solid (dashed) lines labels the even (odd) Landau level fillings.}
\end{figure*}

\subsection{Competition between Valley Zeeman and the locking between spin and orbital order}

\begin{figure*}
\includegraphics[width=0.8\linewidth]{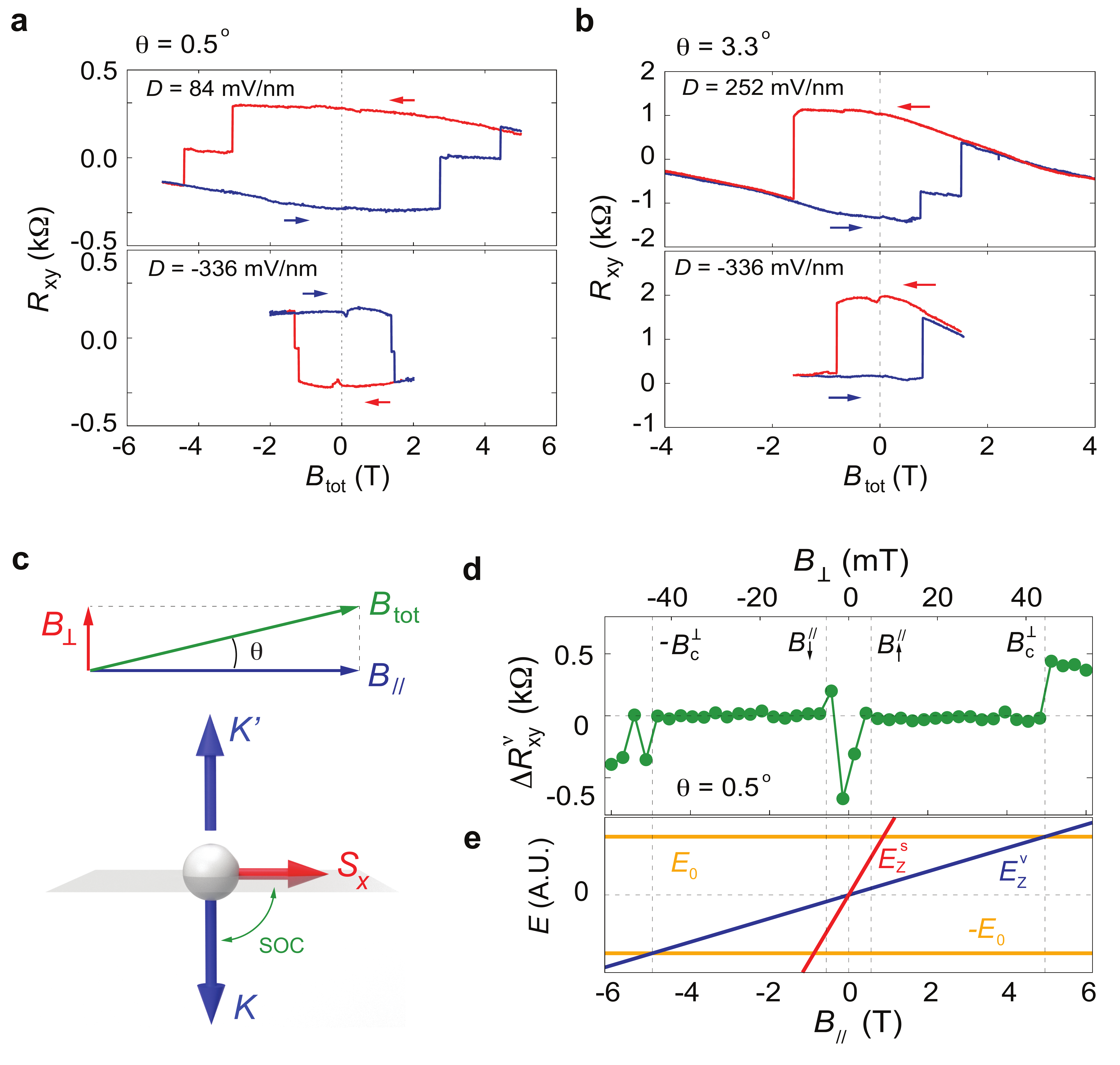}
\caption{\label{fig:competition} {\bf{Competition between different Zeeman coupling mechanism}} (a) Hall resistance as a function of $B$ measured with a tilt angle of $\theta=0.5^{\circ}$. Opposite signs in \Rxy\ are observed at different $D$. The $D$-induced magnetization reversal  indicates that the parallel component of $B$ dominates at this tilt angle. (b) Hysteresis loop as a function of $B$ measured with a tilt angle of $\theta=3.3^{\circ}$. The coercive field becomes smaller at negative $D$, but no magnetization reversal is observed between different $D$, indicating that magnetic order is determined by the perpendicular component of $B$. (c) Schematic showing the effect of tilting the $B$-field. \Bperp\ directly couples to the valley index through valley Zeeman coupling, whereas the influence of \Bpara\ arises from the combination of spin Zeeman and SOC. (d) Residual Hall resistance, \DRxyn, as a function of \Bpara\ (bottom axis) and \Bperp\ (top axis), measured at $\nu=1$, $T= 20$ mK, $D=-400$ mV/nm and with a tilt angle of $\theta = 0.5^{\circ}$. Four transitions are observed in \DRxyn\ at $B$-field values labeled as $B^{\parallel}_{\downarrow}$, $\pm B^{\perp}_{c}$, and $\pm B^{\perp}_{c}$.  
(e) Schematics illustrating the competing energy scales in the presence of a tilted $B$. Spin and valley Zeeman energy, \EZS and \EZV, are linear in $B$, whereas the locking between spin and valley, $E_0$ remains a constant.}
\end{figure*}

The versatile control over the magnetic order allows us to examine the nature of spin-orbital locking through the competition between different coupling mechanisms. For example, \Bperp\ couples to the magnetic order through valley Zeeman, whereas the influence of \Bpara\ arises from the combination of spin Zeeman and SOC.
Fig.~\ref{fig:competition}a-b plot $B$-induced hysteresis loops of $\nu=2$ measured at two different tilt angles. At $\theta = 0.5^{\circ}$, the in-plane component of $B$ and spin Zeeman coupling dominates, $E^{s}_Z > E^{v}_Z$, resulting in a magnetization reversal induced by $D$. On the other hand, the perpendicular component of $B$ and valley Zeeman coupling dominates at $\theta=3.3^{\circ}$, $E^{s}_Z < E^{v}_Z$. As a result, the $D$ dependence in the magnetic order is consistent with Fig.~3c. $\theta$-dependence displayed by the hysteresis loops confirms the different mechanism with which \Bpara\ and \Bperp couple to the magnetic state. 

In the regime where the in-plane component of $B$ and spin Zeeman coupling dominates, $E^{s}_Z > E^{v}_Z$, the magnetic order could provide insight into the relative strength between valley Zeeman and SOC (Fig.~\ref{fig:competition}c).  
Such competition is demonstrated by plotting residual Hall resistance, \DRxyn, as a function of $B$ at a tilt angle $\theta = 0.5^\circ$.
It is worth pointing out that \DRxyn\ changes  sign at $B=0$ (Fig.~2f-g), suggesting that the sign of \DRxyn\ reflects the stable configuration of the magnetic state. 
The fact that \DRxyn\ in Fig.~\ref{fig:competition}d exhibits a series of transitions indicates multiple magnetization reversals. At small $B$, the magnetization is determined by the in-plane component of $B$, with two transitions observed at $B^{\parallel}_{\uparrow}$ and $B^{\parallel}_{\downarrow}$. 
This behavior indicates that the influence of SOC and \EZS\ dominates at small $B$, owing to the small tilt angle $\theta $. With increasing $B$, another set of transitions are observed at $\pm B^{\perp}_{c}$, which points toward the unlocking between spin and valley indices. The value of $B^{\perp}_{c}$ provides an estimate for the energy needed to unlock spin and valley degrees of freedom, $E_0$. Assuming a valley g-factor of $\sim 2$, the value of $E_0 \sim 10 \mu$eV is similar to the intrinsic SOC of graphene, which is several orders of magnitude smaller compared to previous reported values in graphene/\wse\ structures ~\cite{island2019spin}. This could indicate that the valley g-factor is significantly enhanced by strong correlation within the moir\'e flatband, thus increasing valley Zeeman coupling with \Bperp. Alternatively, there could be another energy scale, besides proximity-induced SOC, that is associated with the coupling between the magnetic order and $S_x$. For example, \Bpara-control of the magnetic order relies on a non-zero  $S_x$, which results from the breaking of the $C_3$ symmetry. Therefore, $E_0$ could be associated with the $C_3$ symmetry  breaking in tBLG.

\begin{figure*}
\includegraphics[width=0.7\linewidth]{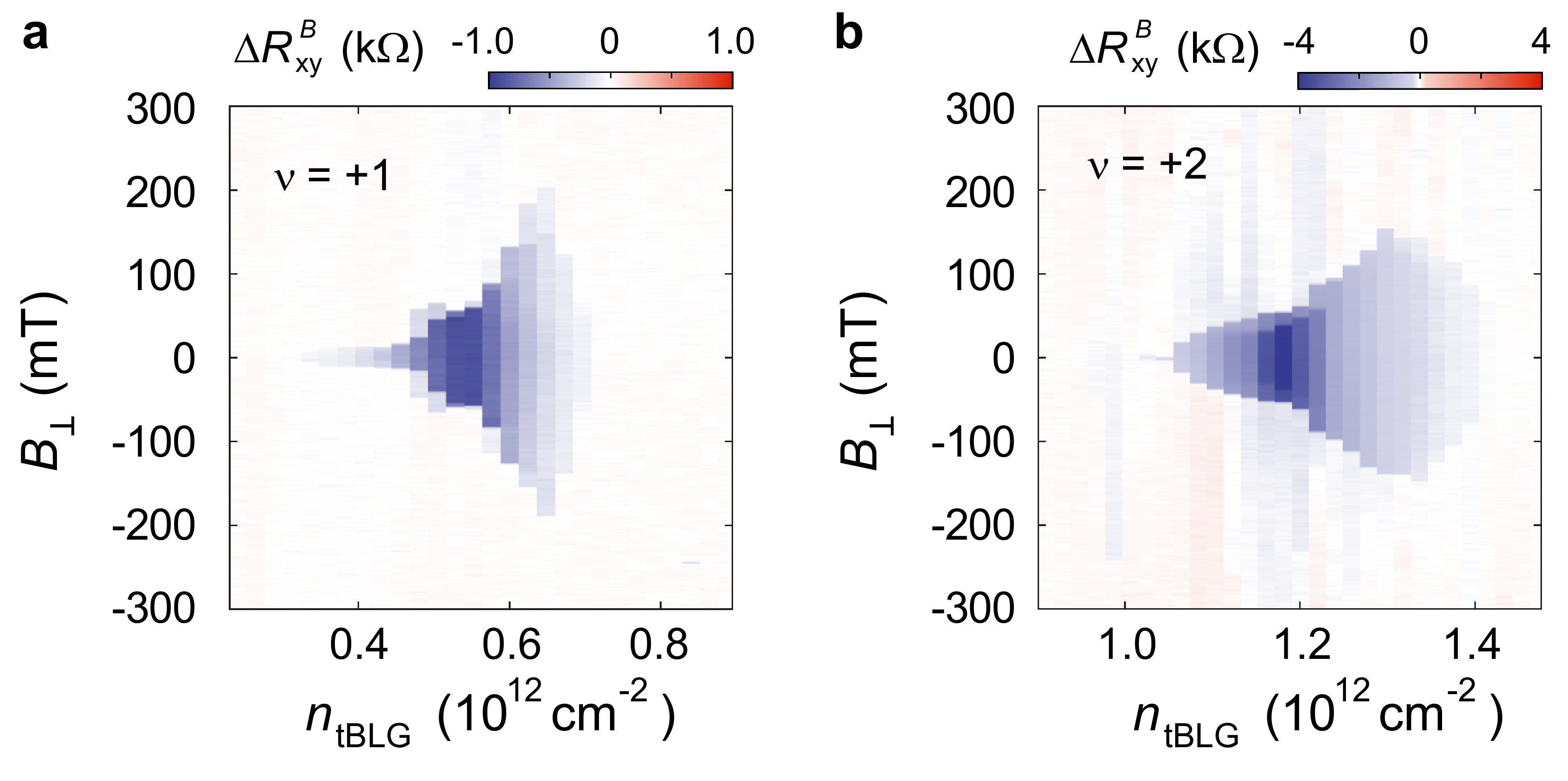}
\caption{\label{fig:Bloop_nmap} {\bf{Residual Hall resistance from $B$-induced hysteresis}} \DRxyb\ as a function of carrier density and \Bperp at (a) $\nu=+1$ and (b) $\nu=+2$. \DRxyb\ is defined as the difference in  \Rxy\ between traces and retraces when the magnetic field is swept back and forth. Each $B$-induced hysteresis loop is measured at a fixed carrier density, with $D=0$ and $T=1.8$~K.}
\end{figure*}

\begin{figure*}
\includegraphics[width=0.7\linewidth]{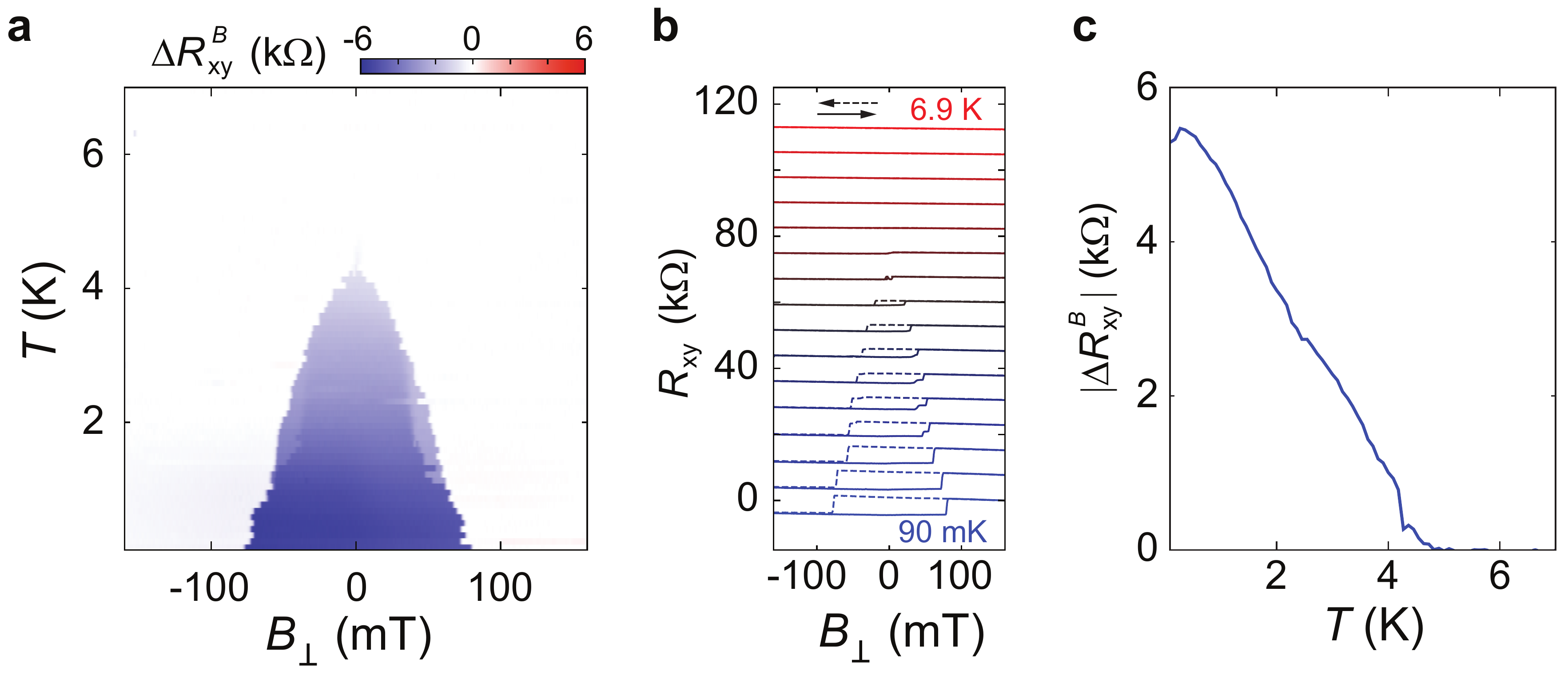}
\caption{\label{fig:Bloop_Tmap} {\bf{Temperature dependence of the $B$-induced hysteresis}} (a) Residual Hall resistance \DRxyb\ in the \Bperp-induced hysteresis loop as a function of temperature and \Bperp. Data taken for $\nu=+2$, at $D=0$ and $n=1.17$\densityunit. (b) Hysteresis loops at different temperatures from (a). Solid (dashed) lines are taken when \Bperp is swept up (down). (c) Line cut from (a) at fixed \Bperp$=0$.}
\end{figure*}


\begin{figure*}
\includegraphics[width=0.5\linewidth]{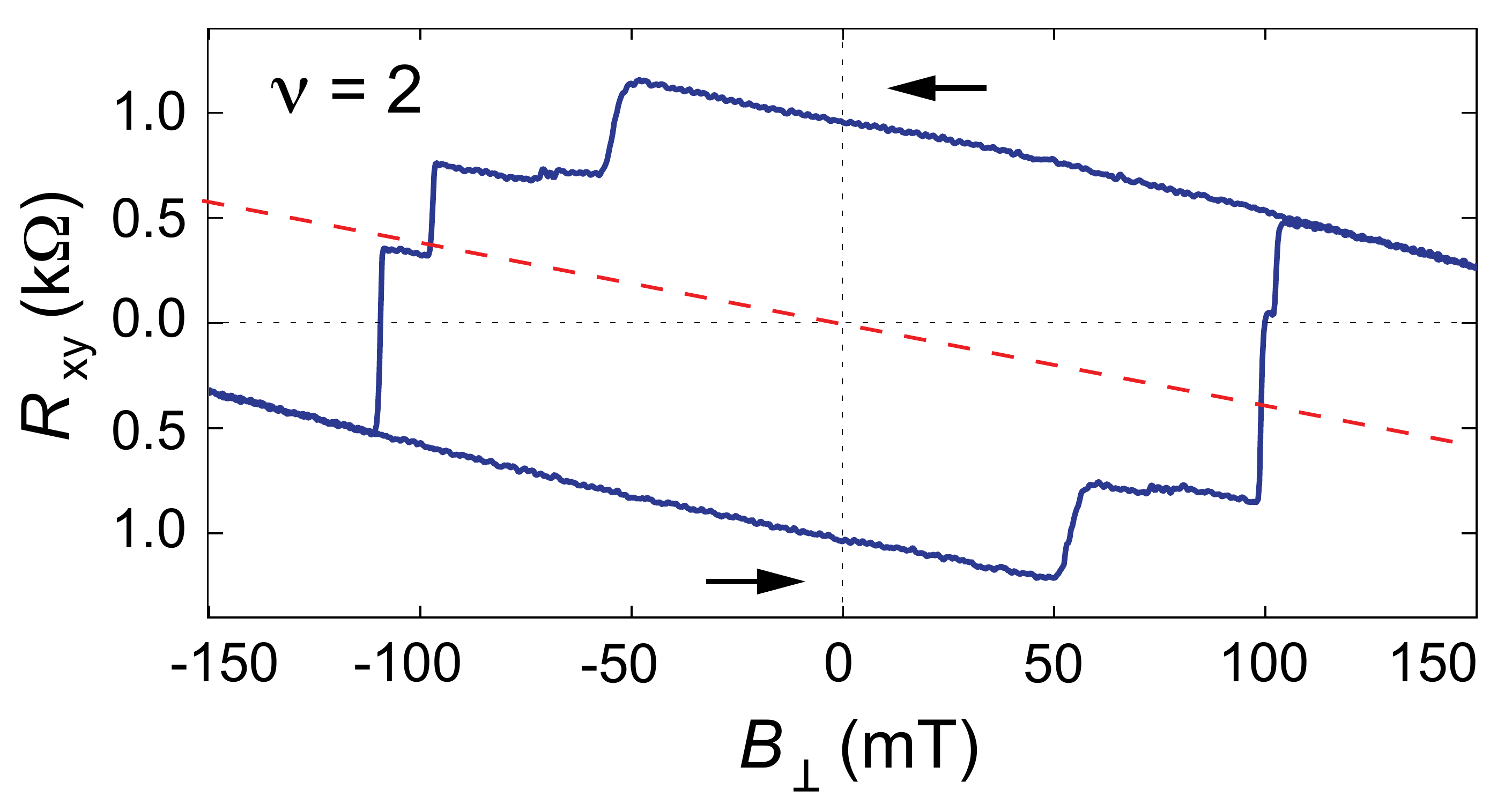}
\caption{\label{fig:Bloopslope} {\bf{Slope of the perpendicular field-induced hysteresis loop}} Hall resistance measured at  $\nu=+2$, $T=1.8$~K, $n_{tBLG}=1.22$\densityunit\ and $D=0$. The red dashed line marks the slope corresponding to the Hall density at this filling (Fig.~\ref{fig:compare}c), showing excellent agreement with the ordinary Hall component of the hysteresis loop. This ordinary Hall component indicates that the Chern gap is not fully developed and the sample bulk is not fully insulating. }
\end{figure*}

\begin{figure*}
\includegraphics[width=0.6\linewidth]{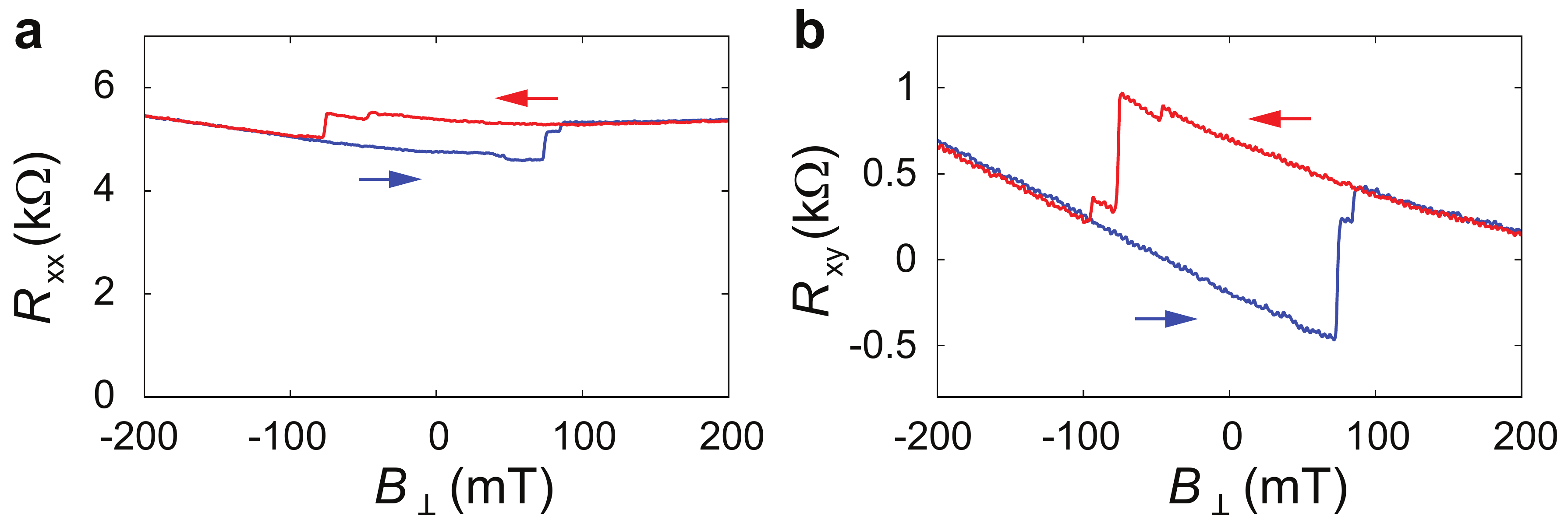}
\caption{\label{fig:RxxRxy} {\bf{Longitudinal and Hall resistance in the hysteresis loop.}} Longitudinal resistance \Rxx\ (a) and Hall resistance \Rxy\ (b) in the perpendicular-magnetic field-induced hysteresis loop at $\nu=+1$, measured at $n=0.55$\densityunit, $D=0$, and $T=2$~K. \Rxx\ is taken with configuration 13-2, 15-16, and \Rxy\ is taken with 13-2, 15-6. The labeling of the configuration follows the description in Fig.\ref{fig:2term}.}
\end{figure*}

\clearpage
\subsection{References and Notes}

{\bibliography{Li_ref}}

\end{widetext}

\end{document}